\documentstyle[12pt,jeep,epsf]{article}

\catcode`\@=11

\let\DOTSI\relax
\def\RIfM@{\relax\ifmmode}
\def\FN@{\futurelet\next}
\newcount\intno@
\def\iint{\DOTSI\intno@\tw@\FN@\ints@}
\def\iiint{\DOTSI\intno@\thr@@\FN@\ints@}
\def\iiiint{\DOTSI\intno@4 \FN@\ints@}
\def\idotsint{\DOTSI\intno@\z@\FN@\ints@}
\def\ints@{\findlimits@\ints@@}
\newif\iflimtoken@
\newif\iflimits@
\def\findlimits@{\limtoken@true\ifx\next\limits\limits@true
 \else\ifx\next\nolimits\limits@false\else
 \limtoken@false\ifx\ilimits@\nolimits\limits@false\else
 \ifinner\limits@false\else\limits@true\fi\fi\fi\fi}
\def\multint@{\int\ifnum\intno@=\z@\intdots@                                
 \else\intkern@\fi                                                          
 \ifnum\intno@>\tw@\int\intkern@\fi                                         
 \ifnum\intno@>\thr@@\int\intkern@\fi                                       
 \int}                                                                      
\def\multintlimits@{\intop\ifnum\intno@=\z@\intdots@\else\intkern@\fi
 \ifnum\intno@>\tw@\intop\intkern@\fi
 \ifnum\intno@>\thr@@\intop\intkern@\fi\intop}
\def\intic@{\mathchoice{\hskip.5em}{\hskip.4em}{\hskip.4em}{\hskip.4em}}
\def\negintic@{\mathchoice
 {\hskip-.5em}{\hskip-.4em}{\hskip-.4em}{\hskip-.4em}}
\def\ints@@{\iflimtoken@                                                    
 \def\ints@@@{\iflimits@\negintic@\mathop{\intic@\multintlimits@}\limits    
  \else\multint@\nolimits\fi                                                
  \eat@}                                                                    
 \else                                                                      
 \def\ints@@@{\iflimits@\negintic@
  \mathop{\intic@\multintlimits@}\limits\else
  \multint@\nolimits\fi}\fi\ints@@@}
\def\intkern@{\mathchoice{\!\!\!}{\!\!}{\!\!}{\!\!}}
\def\plaincdots@{\mathinner{\cdotp\cdotp\cdotp}}
\def\intdots@{\mathchoice{\plaincdots@}
 {{\cdotp}\mkern1.5mu{\cdotp}\mkern1.5mu{\cdotp}}
 {{\cdotp}\mkern1mu{\cdotp}\mkern1mu{\cdotp}}
 {{\cdotp}\mkern1mu{\cdotp}\mkern1mu{\cdotp}}}

\newif\iffirstchoice@
\firstchoice@true
\def\textfonti{\the\textfont\@ne}
\def\textfontii{\the\textfont\tw@}
\def\text{\RIfM@\expandafter\text@\else\expandafter\text@@\fi}
\def\text@@#1{\leavevmode\hbox{#1}}
\def\text@#1{\mathchoice
 {\hbox{\everymath{\displaystyle}\def\textfonti{\the\textfont\@ne}%
  \def\textfontii{\the\textfont\tw@}\textdef@@ T#1}}
 {\hbox{\firstchoice@false
  \everymath{\textstyle}\def\textfonti{\the\textfont\@ne}%
  \def\textfontii{\the\textfont\tw@}\textdef@@ T#1}}
 {\hbox{\firstchoice@false
  \everymath{\scriptstyle}\def\textfonti{\the\scriptfont\@ne}%
  \def\textfontii{\the\scriptfont\tw@}\textdef@@ S\rm#1}}
 {\hbox{\firstchoice@false
  \everymath{\scriptscriptstyle}\def\textfonti
  {\the\scriptscriptfont\@ne}%
  \def\textfontii{\the\scriptscriptfont\tw@}\textdef@@ s\rm#1}}}
\def\textdef@@#1{\textdef@#1\rm\textdef@#1\bf\textdef@#1\sl\textdef@#1\it}
\def\DN@{\def\next@}
\def\eat@#1{}
\def\textdef@#1#2{%
 \DN@{\csname\expandafter\eat@\string#2fam\endcsname}%
 \if S#1\edef#2{\the\scriptfont\next@\relax}%
 \else\if s#1\edef#2{\the\scriptscriptfont\next@\relax}%
 \else\edef#2{\the\textfont\next@\relax}\fi\fi}

\def\Let@{\relax\iffalse{\fi\let\\=\cr\iffalse}\fi}
\def\vspace@{\def\vspace##1{\crcr\noalign{\vskip##1\relax}}}
\def\multilimits@{\bgroup\vspace@\Let@
 \baselineskip\fontdimen10 \scriptfont\tw@
 \advance\baselineskip\fontdimen12 \scriptfont\tw@
 \lineskip\thr@@\fontdimen8 \scriptfont\thr@@
 \lineskiplimit\lineskip
 \vbox\bgroup\ialign\bgroup\hfil$\m@th\scriptstyle{##}$\hfil\crcr}
\def\Sb{_\multilimits@}
\def\endSb{\crcr\egroup\egroup\egroup}
\def\Sp{^\multilimits@}

\newdimen\ex@
\def\rightarrowfill@#1{$#1\m@th\mathord-\mkern-6mu\cleaders
 \hbox{$#1\mkern-2mu\mathord-\mkern-2mu$}\hfill
 \mkern-6mu\mathord\rightarrow$}
\def\leftarrowfill@#1{$#1\m@th\mathord\leftarrow\mkern-6mu\cleaders
 \hbox{$#1\mkern-2mu\mathord-\mkern-2mu$}\hfill\mkern-6mu\mathord-$}
\def\leftrightarrowfill@#1{$#1\m@th\mathord\leftarrow\mkern-6mu\cleaders
 \hbox{$#1\mkern-2mu\mathord-\mkern-2mu$}\hfill
 \mkern-6mu\mathord\rightarrow$}
\def\overrightarrow{\mathpalette\overrightarrow@}
\def\overrightarrow@#1#2{\vbox{\ialign{##\crcr\rightarrowfill@#1\crcr
 \noalign{\kern-\ex@\nointerlineskip}$\m@th\hfil#1#2\hfil$\crcr}}}

\def\overleftarrow{\mathpalette\overleftarrow@}
\def\overleftarrow@#1#2{\vbox{\ialign{##\crcr\leftarrowfill@#1\crcr
 \noalign{\kern-\ex@\nointerlineskip}$\m@th\hfil#1#2\hfil$\crcr}}}
\def\overleftrightarrow{\mathpalette\overleftrightarrow@}
\def\overleftrightarrow@#1#2{\vbox{\ialign{##\crcr\leftrightarrowfill@#1\crcr
 \noalign{\kern-\ex@\nointerlineskip}$\m@th\hfil#1#2\hfil$\crcr}}}
\def\underrightarrow{\mathpalette\underrightarrow@}
\def\underrightarrow@#1#2{\vtop{\ialign{##\crcr$\m@th\hfil#1#2\hfil$\crcr
 \noalign{\nointerlineskip}\rightarrowfill@#1\crcr}}}

\def\underleftarrow{\mathpalette\underleftarrow@}
\def\underleftarrow@#1#2{\vtop{\ialign{##\crcr$\m@th\hfil#1#2\hfil$\crcr
 \noalign{\nointerlineskip}\leftarrowfill@#1\crcr}}}
\def\underleftrightarrow{\mathpalette\underleftrightarrow@}
\def\underleftrightarrow@#1#2{\vtop{\ialign{##\crcr$\m@th\hfil#1#2\hfil$\crcr
 \noalign{\nointerlineskip}\leftrightarrowfill@#1\crcr}}}

\catcode`\@=\active

\def\frac#1#2{{#1 \over #2}}

\def\dfrac#1#2{{\displaystyle {#1 \over #2}}}

\newcount\GRAPHICSTYPE
\def\GRAPHICSPS#1{%
\ifnum\GRAPHICSTYPE=1 language "PS", include "#1"\else%
ps: #1\fi}

\def\graffile#1#2#3#4{\leavevmode\raise -#4 \hbox{%
\raise #3 \hbox{\rule{0.003in}{0.003in}\special{#1}}}%
{\raise -#4 \hbox to #2 {\vrule height#3 width0in depth0in\hfil}}%
}

\def\draftbox#1#2#3#4{\leavevmode\raise -#4 \hbox{\frame{\rlap{\protect\tiny
#1}%
\hbox to #2{\vrule height#3 width0in depth0in\hfil}}}}

\newcount\draft
\def\GRAPHIC#1#2#3#4#5{\ifnum\draft=1 \draftbox{#2}{#3}{#4}{#5}\else%
\graffile{#1}{#3}{#4}{#5}\fi}

\def\addtoLaTeXparams#1{\edef\LaTeXparams{\LaTeXparams #1}}

\def\doFRAMEparams#1{\readFRAMEparams#1\end}
\def\readFRAMEparams#1{%
\ifx#1\end%
\let\next=\relax%
\else%
\ifx#1i%
\dispkind=0%
\fi%
\ifx#1d%
\dispkind=1%
\fi%
\ifx#1f%
\dispkind=2%
\fi%
\ifx#1t%
\addtoLaTeXparams{t}%
\fi%
\ifx#1b%
\addtoLaTeXparams{b}%
\fi%
\ifx#1p%
\addtoLaTeXparams{p}%
\fi%
\ifx#1h%
\addtoLaTeXparams{h}%
\fi%
\let\next=\readFRAMEparams%
\fi%
\next%
}

\def\IFRAME#1#2#3#4#5{\GRAPHIC{#5}{#4}{#1}{#2}{#3}}

\def\DFRAME#1#2#3#4{
  \begin{center}
    \GRAPHIC{#4}{#3}{#1}{#2}{0in}
  \end{center}
}

\def\FFRAME#1#2#3#4#5#6#7{
  \begin{figure}[#1]
    \begin{center}
      \GRAPHIC{#7}{#6}{#2}{#3}{0in}
    \end{center}
    \caption{\label{#5}#4}
  \end{figure}
}

\def\FRAME#1#2#3#4#5#6#7#8{%
\newcount\dispkind%
\def\LaTeXparams{}%
\dispkind=0%
\def\LaTeXparams{}%
\doFRAMEparams{#1}%
\ifnum\dispkind=0%
\IFRAME{#2}{#3}{#4}{#7}{#8}%
\else
  \ifnum\dispkind=1
    \DFRAME{#2}{#3}{#7}{#8}
  \else
    \ifnum\dispkind=2
      \FFRAME{\LaTeXparams}{#2}{#3}{#5}{#6}{#7}{#8}
    \fi
  \fi
\fi
}

\catcode`\@=11

\long\def\QQQ#1#2{}
\def\QTP#1{}
\long\def\QQA#1#2{}

\def\EXPAND#1[#2]#3{}
\def\NOEXPAND#1[#2]#3{}

\def\LaTeXparent#1{}

\@ifundefined{INDEX}{\def\INDEX#1#2{}{}}{}
\@ifundefined{SUBINDEX}{\def\SUBINDEX#1#2#3{}{}{}}{}
\def\initial#1{\bigbreak{\raggedright\large\bf #1}\kern 2pt\penalty3000}

\@ifundefined{abstract}{%
\def\abstract{\if@twocolumn
\section*{Abstract (Not appropriate in this style!)}
\else \small
\begin{center}
{\bf Abstract\vspace{-.5em}\vspace{0pt}}
\end{center}
\quotation
\fi}}{}
\@ifundefined{endabstract}{%
\def\endabstract{\if@twocolumn\else\endquotation\fi}}{}
\@ifundefined{maketitle}{\def\maketitle#1{}}{}
\@ifundefined{affiliation}{\def\affiliation#1{}}{}
\@ifundefined{proof}{}{}
\@ifundefined{newfield}{\def\newfield#1#2{}}{}
\@ifundefined{chapter}{\def\chapter#1{\par(Chapter head:)#1\par }}{}
\@ifundefined{part}{\def\part#1{\par(Part head:)#1\par }}{}
\@ifundefined{section}{\def\part#1{\par(Section head:)#1\par }}{}
\@ifundefined{subsection}{\def\part#1{\par(Subsection head:)#1\par }}{}
\@ifundefined{subsubsection}{\def\part#1{\par(Subsubsection head:)#1\par }}{}
\@ifundefined{paragraph}{\def\part#1{\par(Subsubsubsection head:)#1\par }}{}
\@ifundefined{subparagraph}{\def\part#1{\par(Subsubsubsubsection head:)#1\par
}}{}

\newdimen\theight
\def \Column{%
             \vadjust{\setbox0=\hbox{\scriptsize\quad\quad tcol}%
             \theight=\ht0
             \advance\theight by \dp0    \advance\theight by \lineskip
             \kern -\theight \vbox to \theight{\rightline{\rlap{\box0}}%
             \vss}%
             }}%

\def\qed{\ifhmode\unskip\nobreak\fi\ifmmode\ifinner\else\hskip5\p@\fi\fi
 \hbox{\hskip5\p@\vrule width4\p@ height6\p@ depth1.5\p@\hskip\p@}}
\catcode`@=12 

\newcommand\putfig[3]{
   \vbox{
      \let\picnaturalsize=N
   \def\picsize{#3}
   \def\picfilename{#1}
   \ifx\nopictures Y\else{\ifx\epsfloaded Y\else\input epsf \fi
   \let\epsfloaded=Y
   \centerline{\ifx\picnaturalsize N\epsfxsize \picsize\fi
   \epsfbox{\picfilename}}}\fi
      \vspace{1.0cm}
   {\it #2}
   \vspace{1.5cm}
   }
}

\begin{document}

\pagestyle{plain}
\thispagestyle{empty}
\def\theequation{\thesection.\arabic{equation}}
\setcounter{equation}{0}

\noindent {hep-th/9405156}       \hfill                  {USC-94/HEP-B1}\\
\noindent  {}                    \hfill                   May 1994
\vspace{1.2cm}

\begin{centering}

{\huge Folded Strings\\ Falling Into a Black Hole\footnote
{Research supported in part by the
DOE Grant No. DE-FG03-84ER-40168. }}\\
\vspace{1cm}
{\large Itzhak Bars and J\"urgen Schulze
   \footnote{Supported by the DAAD (Doktorandenstipendium HSPII/AUFE)}\\
\vspace {1cm}
Department of Physics and Astronomy\\
University of Southern California\\
Los Angeles, CA 90089-0484, USA}\\
\end{centering}
\vspace{.25cm}

\begin{abstract}
\vspace {.2cm}

We find all the classical solutions (minimal surfaces) of open or closed
strings in {\it any} two dimensional curved spacetime. As examples we
consider the SL(2,R)/R two dimensional black hole, and any 4D black hole in
the Schwarzschild family, provided the motion is restricted to the
time-radial components. The solutions, which describe longitudinaly
oscillating folded strings (radial oscillations in 4D), must be given in
lattice-like patches of the worldsheet, and a transfer operation analogous
to a transfer matrix determines the future evolution. Then the swallowing of
a string by a black hole is analyzed. We find several new features that are
not shared by particle motions. The most surprizing effect is the tunneling
of the string into the bare singularity region that lies beyond the black
hole that is classically forbidden to particles.
\end{abstract}

\newpage\

\section{Introduction}
\setcounter{equation}{0}

The geodesics of a point particle falling into a black hole are well
understood, but little is known about the ``geodesics'' of strings falling
into a black hole. Since a string has internal degrees of freedom, its
motion in the vicinity of a black hole could have new features that cannot
be guessed by the study of particles. Such a string could be a model for an
extended body with internal degrees of freedom, or particles in a gas with
internal interactions, that are being swallowed by a black hole.
Furthermore, it is possible to imagine very long strings, in definite normal
modes, with one end inside the black hole and the other one outside the
horizon. Can one learn more about the properties of the black hole by
``fishing'' with such strings?

More generally, the motion of strings in curved spacetime is basically not
known. Finding the classical solutions would be helpful for understanding
and interpreting the quantum theory, which is of interest for Cosmology and
Unification ideas. Fortunately, there are some special curved spacetimes
based on gauged WZW models (GWZW) \cite{gwzw} with a single time coordinate
\cite{ibreviews} for which it is possible to obtain the complete set of
classical solutions \cite{ibsfannals}, and in principle the quantum
solutions. One of the additional purposes of our paper is to begin a study
of such models in more detail and provide some interpretation of the
classical solutions.

In this paper we present all the classical string solutions for {\it {any}}
curved $2D$ spacetime. The two dimensions could be interpreted as the
time-radial coordinates of any spherically symmetric four dimensional
spacetime, with the motion occuring at constant angles $d\theta =d\phi =0$.
Examples include the $SL(2,R)/R$ black hole and the Schwarzschild type
curved spaces with general metric of the form $ds^2=-\,dt^2\,f(r)\, +
dr^2/\,f(r)$. These examples allow us to investigate the question of how a
string falls into a black hole. Some work has already been done in this
direction by concentrating on special solutions involving circular strings
\cite{devega}. Instead, we will investigate longitudinal motions of strings,
and we will obtain all the solutions. Our classical solutions describe folded
strings with the folds oscillating against each other while the whole string
is being attracted by the black hole. There are many new physical features,
but the main surprize is the tunneling of the string into the bare
singularity region of spacetime that lies beyond the black hole and which is
forbidden to particles.

We find that naive solutions can be defined only in patches of the
worldsheet while complete solutions are obtained by matching boundary
conditions at the boundaries of the patches. The patches correspond to a
lattice type structure on the worldsheet, with the lattice being different
for each distinct solution. The lattice is dynamically defined by the
initial conditions that are provided by folded string motions in the
asymptotically flat spacetime. The future development of the initial
condition is given in the form of a transfer operation, analogous to a
transfer matrix on the lattice, but with the transfer corresponding to a
period of oscillation of the folds on the string. The patching
procedure is applied explicitly and a complete solution is obtained and
physically analyzed.

A word of caution: In this paper the classical theory of strings in curved
spacetime is studied on its own merit. However, when the string reaches the
black hole the correct physics may require a full quantum treatment as well
as the back reaction of the matter on the black hole. These issues which may
be important are not studied here. The effects found in this paper are
therefore only tentative, however we feel that they signal some new
phenomena.

The paper is organized as follows. In section 2 we present the general
string solution in any $2D$ metric (which may be the $(r,t)$ restriction of
a $4D$ metric). In section 3 the same solution is obtained with special
methods appropriate for gauged WZW models. In section 4 a cell decomposition
of the worldsheet is introduced using the yo-yo solution in flat spacetime
as an example. In section 5.1 the boundary matching procedure is dicussed
for any metric. In section 5.2 the procedure is applied explicitly on the
$2D$ black hole based on $SL(2,R)/R$ to obtain a complete solution and
derive the transfer operation. In section 5.3
the transfer matrix is constructed and an shown that
a discrete version of the minimal area is one of its invariants. In
section 5.4 general solutions are described. Finally the physics is
discussed in section 6 where the motion is described with the help of
figures and the new features are emphasized.

\section{String solution in any 2D metric}
\setcounter{equation}{0}

Let us consider a string $x^\mu (\tau ,\sigma )$ propagating in a curved
spacetime manifold. The string Lagrangian in the conformal gauge is given by
\begin{equation}
\label{stringl}L=\partial _{+}x^\mu \partial _{-}x^\nu \,G_{\mu \nu }(x).
\end{equation}
where $\partial _{\pm }=(\partial _\tau \pm \partial _\sigma )/\sqrt{2}.$
The equations of motion are just the geodesic equations for a string. In
addition, these are supplemented with constraints that come from the
reparametrization invariance of the theory (zero energy-momentum tensor)
\begin{equation}
\label{stringconstraints}G_{\mu \nu }\,\partial _{+}x^\mu \partial _{+}x^\nu
=0=G_{\mu \nu }\,\partial _{-}x^\mu \partial _{-}x^\nu .
\end{equation}
The general string Lagrangian may include an antisymmetric tensor, a
dilaton, tachyon or other string condensates. In 2D the antisymmetric tensor
can be eliminated since it corresponds to a total divergence. All of the
other condensates are generated by quantum corrections. However, our purpose
here is to study the classical limit of the theory in the absence of all
these fields. Therefore we will define the classical propagation of strings
in curved spacetime through the classical equations of motion and
constraints given above.

We may always redefine the string field by a coordinate reparametrization
$x^\mu (\tau ,\sigma )=x^\mu (y(\tau ,\sigma )).$ Substituting it into the
Lagrangian we see that this gives a new target space metric in the
coordinates $y^\mu (\tau ,\sigma )$ that is related to the old one by a
general coordinate transformation $(\partial x^\lambda /\partial y^\mu
)\,(\partial x^\sigma /\partial y^\nu )\,G_{\lambda \sigma }(y)$. In $2D$ it
is always possible to choose $x^\mu (y)$ such that the new metric is
conformal. Therefore, we might as well begin our analysis with the conformal
metric in target space $G_{\mu \nu }=G(x)\eta _{\mu \nu }.$

It is convenient to define the lightcone coordinates
\begin{equation}
\label{lightcone}
u=\frac 1{\sqrt{2}}(x^0+x^1),\quad v=\frac 1{\sqrt{2}}(x^0-x^1)
\end{equation}
Then the equations of motion and constraints take the form
\begin{equation}
\label{stringeqs}
\begin{array}{c}
\partial _{+}(G\,\,\partial _{-}u)+\partial _{-}(G\,\,\partial _{+}u)=
\frac{\partial G}{\partial v}(\partial _{+}u\partial _{-}v+\partial
_{+}v\partial _{-}u) \\ \partial _{+}(G\,\,\partial _{-}v)+\partial
_{-}(G\,\,\partial _{+}v)=
\frac{\partial G}{\partial u}(\partial _{+}u\partial _{-}v+\partial
_{+}v\partial _{-}u) \\ \partial _{+}u\partial _{+}v=0=\partial
_{-}u\partial _{-}v\,\,\,.
\end{array}
\end{equation}
There are four classes of solutions that may be verified explicitly by
substitution into the differential equation:
\begin{equation}
\label{fourr}
\begin{array}{ll}
A:\qquad u=u(\sigma ^{+}), & \quad v=v(\sigma ^{-}) \\
B:\qquad u=\bar u(\sigma ^{-}), & \quad v=\bar v(\sigma _{\ }^{+}) \\
C:\qquad u=c_1, & \quad v=v(a(\sigma ^{-}),b(\sigma ^{+})) \\
D:\qquad u=u(\bar a(\sigma ^{+}),\bar b(\sigma ^{-})), & \quad v=c_2\,\,\,,
\end{array}
\end{equation}
where the functions $u(\sigma ^{+}),v(\sigma ^{-}),\bar u(\sigma ^{-}),\bar
v(\sigma ^{+}),a(\sigma ^{-}),b(\sigma ^{+}),\bar a(\sigma ^{+}),\bar
b(\sigma ^{-})$ are arbitrary and $c_1,c_2$ are constants. Solutions $A,B$
are present for any metric and do not depend on $G$, but solutions $C,D$
depend on the metric as follows
\begin{equation}
\label{foura}
\begin{array}{c}
C:\quad u=c_1\quad \quad \int^{v(\sigma ^{+},\sigma ^{-})}dv^{\prime
}G(c_1,v^{\prime })=\alpha (\sigma ^{+})+\beta (\sigma ^{-})\,\,\,, \\
\\
D:\quad v=c_{2,}\quad \quad \int^{u(\sigma ^{+},\sigma ^{-})}du^{\prime
}G(u^{\prime },c_2)=\tilde \alpha (\sigma ^{+})+\tilde \beta (\sigma ^{-}),
\end{array}
\end{equation}
where the integration is performed at constant $u=c_1$ for solution $C,$ and
at constant $v=c_2$ for solution $D.$ The functions $\alpha ,\beta ,\tilde
\alpha ,\tilde \beta $ are arbitrary. The arbitrary functions $a(\sigma
^{-}),b(\sigma ^{+}),\bar a(\sigma ^{+}),\bar b(\sigma ^{-})$ that appear in
(\ref{fourr}) are these or their reparametrizations that may be more
convenient. Taking a derivative of these integrals with respect to $\sigma
^{\pm }$ gives relations that solve the equations of motion.

For a given explicit metric $G(uv)$ one obtains a metric dependent solution
$C$ or $D$ after performing these integrals. For example, for the flat metric
$G=1$ one has the obvious solution $u(\sigma ^{+},\sigma ^{-})=\alpha
(\sigma ^{+})+\beta (\sigma ^{-}),$ or $v(\sigma ^{+},\sigma ^{-})=\tilde
\alpha (\sigma ^{+})+\tilde \beta (\sigma ^{-}).\,\,$. Similarly, for the
$2D$ black hole one has $G=(1-uv)^{-1},$ and after performing the integrals
and doing some reparametrizations one obtains the solutions
\begin{equation}
\label{fourbh}
\begin{array}{c}
\begin{array}{ll}
C:\qquad u=c_1, & \quad v=\frac 1{c_1}[1-a(\sigma ^{-})\,\,b(\sigma ^{+})],
\\
&  \\
D:\qquad v=c_2\,, & \quad u=\frac 1{c_2}[1-\bar a(\sigma ^{+})\,\,\bar
b(\sigma ^{-})].
\end{array}
\\
\end{array}
\end{equation}

The solutions for the 2D black hole metric were obtained with different
methods by various groups at different stages. The first attempts were in
the context of the 2D black hole \cite{bastony}, which later led to the
general classical solution for any gauged WZW model \cite{ibsfannals} (see
also next section). This approach, which gives the complete set of
solutions, leads directly to the set above. In an independent approach, the
string equations for the $2D$ black hole were also solved directly in \cite
{devega2}. On the basis of their solutions, the authors of \cite{devega2}
came to the conclusion that there are only particle solutions (i.e. only
collapsed strings) when the constraints (\ref{stringconstraints}) are taken
into account. However, as we will discuss below there are an infinite number
of stringy solutions in the $2D$ black hole background. Our conclusions are
quite different because we take into account folded strings. In fact, we
advertized some time ago in \cite{berkeley}, that, in the black hole curved
spacetime, we obtained the analogs of the folded string solutions that were
discussed by BBHP \cite{bbhp} in flat spacetime.

Here we also give the string solutions for the Schwarzschild-like metrics of
the form
\begin{equation}
\label{kruskala}ds^2=f(r)\,\,dt^2-\frac 1{f(r)}dr^2=G(uv)\,\,dudv
\end{equation}
Examples are $f(r)=1-1/r\,$ for the Schwarzschild black hole, and $f(r)=
1-1/r+q^2/r^2$ for the charged black hole, etc.. The transformation to
Kruskal coodinates gives
\begin{equation}\label{kruskalb}
uv=-\exp \left( 2\int^r\frac{dr^{\prime }}{f(r^{\prime })} \right) \quad ,
\quad G(uv)=-f(r)\,\exp \left( -2\int^r\frac{dr^{\prime }}
{f(r^{\prime })}\right) ,
\end{equation}
and $u/v=\exp (2t).$ In particular, for the Schwarzschild black hole case
$uv=(1-r)\exp (r)$ and $G=\exp (-r)/r.$ Once written in Kruskal coordinates
we analytically continue to the global space that includes all spacetime
regions, including those outside the horizon, inside the horizon and the
naked singularity region. Furthermore, it may be necessary to take
multicovers of the $(u,v)$ space in order to have a geodesically complete
manifold.

Having identified the conformal factor, the $C,D$ type string solutions are
obtained by performing the integrals in (\ref{foura}) as follows
\begin{equation}
\label{schwstringa}\int^{u(\sigma ^{+},\sigma ^{-})}du^{\prime }G(u^{\prime
}c_2)=\int^{r(uc_2)}dr^{\prime }\frac{\partial u^{\prime }}{\partial
r^{\prime }}G(r^{\prime })=\frac 1{c_2}r(uc_2)=\tilde \alpha (\sigma
^{+})+\tilde \beta (\sigma ^{-})
\end{equation}
where we have used, $G(r)[\partial (uc_2)/\partial r]=1,$ so that the
integral is easily done for any $f(r).$ Replacing this result back in the
relation between $uv$ and $r$ we get the string solutions, e.g. for the
Schwarzschild metric,
\begin{equation}
\label{schwstringb}
\begin{array}{c}
C:\quad u=c_1\quad \quad \quad v(a(\sigma ^{-}),\,\,b(\sigma ^{+})\,)=\frac
1{c_1}a(\sigma ^{-})\,\,b(\sigma ^{+})\,\,[1-\ln (a(\sigma ^{-})\,\,b(\sigma
^{+}))]
\text{ } \\  \\
D:\quad v=c_2\quad \quad \quad u(\bar a(\sigma ^{+}),\,\,\bar b(\sigma
^{-}))=\frac 1{c_2}\bar a(\sigma ^{+})\,\,\bar b(\sigma ^{-})\,\,[1-\ln
(\bar a(\sigma ^{+})\,\,\bar b(\sigma ^{-}))] \\
\end{array}
\end{equation}
where we reparametrized the arbitrary functions to more convenient forms.

The solutions (\ref{fourr}-\ref{schwstringb}) are not yet string solutions.
A string solution must be periodic in $\sigma $, and furthermore for a valid
classical description the global time coordinate $T(\tau ,\sigma )=(u+v)/
\sqrt{2}$ must be an {\it increasing} function of $\tau $ for any $\sigma$.
Backtracking solutions must be excluded from the classical theory (at least
in the $uv<1$ region), just as they are excluded for the free relativistic
point particle, or the free string in flat spacetime \footnote{For the
relativistic point particle one usually chooses the timelike gauge
$x^0(\tau )=p^0\tau ,$ with positive energy $p^0=\sqrt{\vec p^2+m^2},$ in
order to insure a time coordinate that increases with proper time. Of
course, there are also negative energy solutions to the constraint $p_\mu
^2=m^2$ that describe anti-particles. By making the gauge choice to be
monotonicaly increasing (or monotonically decreasing) in $\tau $ one
excludes the possibility of worldlines that backtrack on themselves. This is
required on the basis of physical description in the classical theory so
that particles and anti-particles do not appear in the same classical
solution. The same remarks apply to strings, in the sense that worldsheets
are not allowed to backtrack in the physical time direction. To achieve
this, the physical time coordinate must always increase monotonically as a
function of the proper time $\tau$. In this way anti-strings are excluded
from appearing in the same solution with strings in the classical
description in flat or curved spacetime.}. These two requirements turn out
to be incompatible for any of the solutions above, if a single solution is
required to be valid everywhere on the worldsheet\footnote{The $C,D$
solutions are actually valid in the entire worldsheet but, if
taken on their own, they describe a massless particle, not a string. This
can be seen, e.g. for the $C$ solution, by eliminating $\tau $ in favor of
the target time coordinate from $T=(c_1+v(\tau ,\sigma ))/\sqrt{2},$ and
substituting into the space coordinate $X=(c_1-v(\tau ,\sigma ))/\sqrt{2},$
in order to get the target space relation between position and time
$X=(2c_1-T)/\sqrt{2}.$ This last relation describes the motion of a free
massless particle, not a string, since the $\sigma $ dependence has
disappeared.}. Therefore we have to consider their validity in patches of
the worldsheet and then learn how to combine them through boundary matching
conditions that turn them into valid string solutions. For these reasons we
find it necesary to split the worldsheet into cells, as defined in section 4.

\section{Classical strings and the gauged WZW model}
\setcounter{equation}{0}

{}From the above discussion it is not evident that the solutions presented
form a complete set. For this reason we would like to use the curved
spacetime approach based on $G/H$ gauged WZW models \cite{ibreviews}, in
particular for $SL(2,R)/R$ \cite{BN}, that describes a stringy 2D black hole
\cite{WIT}. In these models we can discuss the completeness and other issues
with more convincing methods.

In previous work \cite{ibsfglobal} it was shown that for a $G/H$ model the
{\it global} sigma model variables (i.e. string coordinates) are those
combinations of group parameters in $G$ that are invariant under the gauge
group $H$. Therefore, we have argued that a classical solution to the
{\it gauge invariant} string coordinates can be obtained by analyzing the
original gauged WZW classical equations in any convenient gauge. Using this
technique, the general geodesics of a particle in the global manifold were
given for any gauged WZW model. Similarly, all the string solutions were
also given \cite{ibsfannals}, but in the string case, the details of
building the gauge invariants were left out of the explicit discussion.

The case of the 2D black hole is very simple since one can identify the
global gauge invariant string coordinates as the Kruskal coordinates $(u,v)$
that parametrize the $SL(2,R)$ group element as
\begin{equation}
\label{det}
g=\left( \begin{array}{cc} \ \ u & a \\ -b & v \end{array} \right)
\qquad uv+ab=1,
\end{equation}
where we gauge a vector-like subgroup that acts as $g\rightarrow \Lambda
g\Lambda ^{-1}$, with
\begin{equation}
\Lambda =\left(
\begin{array}{cc}
\kappa & 0 \\
0 & \kappa ^{-1}
\end{array}
\right) .
\end{equation}
Evidently $(u,v)$ are invariants under the action of the gauge group.
Therefore classical solutions obtained for $(u,v)$ by using any gauge, will
be valid as solutions in any other gauge. As discovered in \cite{ibsfglobal},
as long as one uses the gauge invariant global coordinates one captures
all dual patches of the geometry, and therefore it is immaterial if one
gauges a vector or an axial subgroup.

The classical equations of motion for the $G/H$ GWZW model are \cite
{ibsfannals}
\begin{equation}
\label{eqs1}(D_{+}gg^{-1})_H=0=(g^{-1}D_{-}g)_H,\qquad F_{+-}=0,\qquad
D_{-}(D_{+}gg^{-1})_{G/H}=0.
\end{equation}
Furthermore, the conformal (Virasoro) constraints require a vanishing stress
tensor

\begin{equation}
\label{con}Tr\left[ (D_{+}gg^{-1})_{G/H}\right] ^2=0=Tr\left[
(g^{-1}D_{-}g)_{G/H}\right] ^2.
\end{equation}
For the $SL(2,R)/R$ case these equations may be explicitly given in the form
\begin{equation}
\label{eqs}
\begin{array}{c}
\begin{array}{cc}
(v\partial _{+}u+b\partial _{+}a)-2abA_{+}=0 & (u\partial _{-}v+b\partial
_{-}a)-2abA_{-}=0 \\
(\partial _{-}+2A_{-})(v\overleftrightarrow{\partial _{+}}b+2vbA_{+})=0\quad
\ \  & \quad \ (\partial _{-}-2A_{-})(u\overleftrightarrow{\partial _{+}}
a-2uaA_{+})=0
\end{array}
\\
\partial _{+}A_{-}-\partial _{-}A_{+}=0 \\
\begin{array}{c}
(v
\overleftrightarrow{\partial _{-}}a-2vaA_{-})(u\overleftrightarrow{\partial
_{-}}b+2ubA_{-})=0 \\ (u\overleftrightarrow{\partial _{+}}a-2uaA_{+})(v
\overleftrightarrow{\partial _{+}}b+2vbA_{+})=0
\end{array}
\end{array}
\end{equation}
They may be rewritten in several other forms by using the determinant
condition in (\ref{det}) and its derivative $\partial _{\pm }(uv+ab)=0.$ To
solve these equations, one approach is to choose the gauge $a=b$ (or $a=-b$
), solve the algebraic equations for $A_{\pm }$ and recognize the remaining
equations for $(u,v)$ as the equations (\ref{stringeqs}) of the string
coordinates moving in the black hole. The solution to these equations are
given in (\ref{fourr}, \ref{fourbh}). Another approach advocated in \cite
{ibsfannals} is to choose the axial gauge $A_{+}=0\,,$ which leads through
the equations of motion to also $A_{-}=0$, and then analyse the remaining
equations for the solution of $(u,v,a,b)$. Since the string coordinates $u,v$
are gauge invariant the solution obtained in one gauge is as good as the
solution obtained in any other gauge.

In the axial gauge $A_{+}=0$ the solution can be written as follows. First,
the group element $g(\tau ,\sigma )$ is given as the product of left and
right moving group elements $g_L(\tau +\sigma )$ and $g_R(\tau -\sigma )$ as
\begin{equation}
\label{g}g=g_Lg_R^{-1}
\end{equation}

\begin{equation}
\label{uv}\left(
\begin{array}{cc}
\,\,\,u & a \\
-b & v
\end{array}
\right) =\left(
\begin{array}{cc}
\,\,\,\,u_L & a_L \\
-b_L & v_L
\end{array}
\right) \left(
\begin{array}{cc}
v_R & -a_R \\
b_R & \,\,u_R
\end{array}
\right)
\end{equation}
therefore
\begin{equation}
\label{uuvv}
\begin{array}{l}
u=u_Lv_R+a_Lb_R\qquad a=a_Lu_R-u_La_R \\
b=b_Lv_R-v_Lb_R\,\qquad v=v_Lu_R+b_La_R\,\,,
\end{array}
\end{equation}
where there are determinant constraints $u_Lv_L+a_Lb_L=1=u_Rv_R+a_Rb_R.$
Then, the left and right moving currents take the form
\begin{equation}
\label{current}\partial _{+}g_Lg_L^{-1}=\left(
\begin{array}{cc}
v_L\partial _{+}u_L+b_L\partial _{+}a_L & u_L\partial _{+}a_L-a_L\partial
_{+}u_L \\
b_L\partial _{+}v_L-v_L\partial _{+}b_L & u_L\partial _{+}v_L+a_L\partial
_{+}b_L
\end{array}
\right) ,
\end{equation}
and similarly for $g_R$, with the derivative replaced by $\partial _{-}$.
Note that after taking into account the determinant conditions $%
a_Lb_L+u_Lv_L=1$, and similarly for right movers, the diagonal entries for
the currents are proportional to the Pauli matrix $\sigma _3.$ The equations
above reduce to the following: (i) the currents that belong to the subgroup $%
H$ (i.e. proportional to $\sigma _3$) vanish
\begin{equation}
\label{proj}(\partial _{+}g_Lg_L^{-1})_H=v_L\partial _{+}u_L+b_L\partial
_{+}a_L=-u_L\partial _{+}v_L-a_L\partial _{+}b_L=0
\end{equation}
and (ii) the conformal constraint becomes
\begin{equation}
\label{constraint}(u_L\overleftrightarrow{\partial _{+}}a_L)\ (b_L
\overleftrightarrow{\partial _{+}}v_L)=0\ ,
\end{equation}
and similarly for right movers. Thus, the equations for left and right
movers separate. A common solution of (\ref{proj}, \ref{constraint}) is
necessarily either $\partial _{+}u_L=\partial _{+}a_L=0$ or $\partial
_{+}v_L=\partial _{+}b_L=0,$ and similarly for the right movers. Therefore
the general solution for left and right movers is of the form
\begin{equation}
\label{form}
\begin{array}{ccc}
g_{L1}=\left(
\begin{array}{cc}
u_{0L} & a_{0L} \\
-\frac 1{a_{0L}}+\frac{u_{0L}}{a_{0L}}v_L(\sigma ^{+}) & v_L(\sigma ^{+})
\end{array}
\right) & \text{or} & g_{L2}=\left(
\begin{array}{cc}
u_L(\sigma ^{+}) & \frac 1{b_{0L}}-
\frac{v_{0L}}{b_{0L}}u_L(\sigma ^{+)} \\ -b_{0L} & v_{0L}
\end{array}
\right) \\
&  &  \\
g_{R1}^{-1}=\left(
\begin{array}{cc}
v_R(\sigma ^{-}) & -a_{0R} \\
\frac 1{a_{0R}}-\frac{u_{0R}}{a_{0R}}v_R(\sigma ^{-}) & u_{0R}
\end{array}
\right) & \text{or} & g_{R2}^{-1}=\left(
\begin{array}{cc}
v_{0R} & -\frac 1{b_{0R}}+
\frac{v_{0R}}{b_{0R}}u_R(\sigma ^{-}) \\ b_{0R} & u_R(\sigma ^{-})
\end{array}
\right)
\end{array}
\end{equation}
Thus, four arbitrary functions $(u_L(\sigma ^{+}),v_L(\sigma
^{+}),u_R(\sigma ^{-}),v_R(\sigma ^{-}))$ together with initial condition
constants $(u_{0L},v_{0L},a_{0L},b_{0L})$ and $(u_{0R},v_{0R},a_{0R},b_{0R})$
parametrize all solutions.

Let us consider some interval of $(\sigma ^{+},\sigma ^{-})$ and build $%
g(\tau ,\sigma )$ by multiplying these solutions. There are four
possibilities in this interval: $\
g_A=g_{L2}g_{R2}^{-1},\;g_B=g_{L1}g_{R1}^{-1},\ g_C=g_{L1}g_{R2}^{-1},\
g_D=g_{L2}g_{R1}^{-1}$. After some rewriting these take the form
\begin{equation}
\label{matrixsolution}
\begin{array}{l}
g_A(\tau ,\sigma )=\left(
\begin{array}{ll}
\ \ u(\sigma ^{+})\quad & \frac 1{b_0}[1-u(\sigma ^{+})v(\sigma ^{-})] \\
-b_0\quad & v(\sigma ^{-})
\end{array}
\right) \\
\\
g_B(\tau ,\sigma )=\left(
\begin{array}{ll}
\bar u(\sigma ^{-}) & \quad a_0 \\
\frac 1{a_0}[-1+\bar u(\sigma ^{-})\bar v(\sigma ^{+})] & \quad \bar
v(\sigma ^{+})
\end{array}
\right) \\
\\
g_C(\tau ,\sigma )=\left(
\begin{array}{ll}
\quad c_1\quad & a(\sigma ^{-}) \\
-b(\sigma ^{+})\qquad & \frac 1{c_1}[1-a(\sigma ^{-})b(\sigma ^{+})]
\end{array}
\right) \\
\\
g_D(\tau ,\sigma )=\left(
\begin{array}{ll}
\frac 1{c_2}[1-\bar a(\sigma ^{+})\bar b(\sigma ^{-})] & \quad \quad \bar
a(\sigma ^{+}) \\
-\bar b(\sigma ^{-}) & \qquad c_2
\end{array}
\right)
\end{array}
\end{equation}
The diagonal entries correspond to the $A,B,C,D$ solutions given in (\ref
{fourr}, \ref{fourbh}). This method which gives a deeper insight also
provides a proof that we have a complete set of solutions.

However, as already mentioned at the end of the previous section, these are
not acceptable string solutions as they stand. We must take the solutions to
be valid only in the patches for which $T$ increases, match boundary
conditions between these patches and impose periodicity. To illustrate this
we will first reexamine flat spacetime solutions.

\section{Cells on the worldsheet}
\setcounter{equation}{0}

As a guide for acceptable curved spacetime string solutions we first
consider the worldsheet structure for flat spacetime solutions. Since in the
asymptotically flat region of any curved spacetime we must have these
solutions as boundary conditions, this analysis will lead directly to the $%
A,B,C,D$ patterns that we are seeking on the worldsheet.

The general flat spacetime solution with a {\it fixed center of mass} was
given in BBHP \cite{bbhp}
\begin{equation}
\label{olde}x^0(\tau ,\sigma )=\frac ML\tau ,\quad \quad x^1(\tau ,\sigma
)=q+\frac M{2L}\,[f(\tau +\sigma )+g(\tau -\sigma )],
\end{equation}
where the slopes of the functions $f^{\prime }=\pm 1,\;$ $g^{\prime }=\pm 1$
may jump from one value to the other in patches of $\tau \pm \sigma $, but
the functions $f,g$ are continuous, and they are {\it {periodic}}. $\frac ML$
is the mass density of the string. The physical length of the string is
proportional to its mass, which is $M$ for the open string and $2M$ for the
closed string. The simplest yo-yo solution is
\begin{equation}
\label{oldee}x^0(\tau ,\sigma )=\frac ML\tau ,\quad \quad x^1(\tau ,\sigma
)=q+\frac M{2L}\,[\ |\tau +\sigma |_{per}+\ |\tau -\sigma |_{per}]\ ,
\end{equation}
where the absolute value is repeated periodically with a period of $\Delta
\sigma =L=4.$

These solutions manifestly satisfy the physical requirements of periodicity
in $\sigma $ and forward propagation in $\tau $ (i.e. $x^0(\tau ,\sigma )$
monotonically increases with $\tau ,$ for all $\sigma $). Our aim in this
section is to clarify how these properties fit together with the $A,B,C,D$
solutions that are valid only in patches. For this purpose it is convenient
to rewrite (\ref{oldee}) in the form
\begin{equation}
\label{old}
\begin{array}{c}
u(\sigma ^{+},\sigma ^{-})=u_0+
\frac{p^{+}}2\left[ (\sigma ^{+}+\left| \sigma ^{+}\right| _{per})+(\sigma
^{-}+\left| \sigma ^{-}\right| _{per})\right] \\ v(\sigma ^{+},\sigma
^{-})=v_0+\frac{p^{-}}2\left[ (\sigma ^{+}-\left| \sigma ^{+}\right|
_{per})+(\sigma ^{-}-\left| \sigma ^{-}\right| _{per})\right]
\end{array}
\end{equation}
For $p^{+}=p^{-}=\sqrt{2}M/L$ the two expressions agree, however (\ref{old})
is slightly more general in that it also includes the motion of the center
of mass when $p^{+}\neq p^{-}.$ The periodic absolute value function $\left|
\sigma ^{+}\right| _{per}\,$ is given by ($\pm \sigma ^{+}-const.)$ where
the value of the constant and the $\pm $ signs depend on regular intervals
in $\sigma ^{+},$ and similarly for $\sigma ^{-}.$ Therefore, the two sets
of signs lead to four possible types of intervals in which the expressions
for $u(\sigma ^{+},\sigma ^{-})$ and $v(\sigma ^{+},\sigma ^{-})$ take four
different looking forms. These forms are precisely of the $A,B,C,D$ types,
specialized to flat spacetime
\begin{equation}
\label{fourflat}
\begin{array}{rl}
(+,-)\qquad A: & u=u_A+p_A^{+}\sigma ^{+},\quad v=v_A+p_A^{-}\sigma ^{-} \\
(-,+)\qquad B: & u=u_B+p_B^{+}\sigma ^{-},\quad v=v_B+p_B^{-}\sigma ^{+} \\
(-,-)\qquad C: & u=u_C,\quad v=v_C+p_C^{+}\sigma ^{+}+p_C^{-}\sigma ^{-} \\
(+,+)\qquad D: & v=v_C,\quad u=u_D+p_D^{+}\sigma ^{+}+p_D^{-}\sigma ^{-} \\
&
\end{array}
\end{equation}
where the momenta are all related $p_A^{\pm }=p_B^{\pm }=p_C^{\pm }=p_D^{\pm
}=p^{\pm }$.

To be more precise, we need to define a cell decomposition of the worldsheet
and give the value of $u,v$ in each cell as a map from the worldsheet to
target spacetime. The patterns of the $A,B,C,D$ solutions assigned to these
cells is the key for obtaining the physical solutions that satisfy the
periodicity and forward propagation requirements. Therefore, we now turn to
the definition of the cells and the patterns.

The solution given in (\ref{old}) decomposes the worldsheet in the following
way. The worldsheet
labelled by $\sigma $ horizontally and by $\tau $ vertically is sliced by
$45^o$ lines that form a light-cone lattice with equal spacings in
$\sigma^\pm $. The crosses in figure (\ref{sol}) below represent the
corners of the cells on the worldsheet.

\begin{equation}
\label{sol}
\begin{array}{ccccccccc}
\nwarrow & \vdots &  &  &  &  &  & \vdots & \nearrow \\
\cdots & \times & B(1,3) & \times & A(2,2) & \times & B(3,1) & \times &
\cdots \\
&  &  &  &  &  &  &  &  \\
& D(0,3) & \times & C(1,2) & \times & D(2,1) & \times & C(3,0) &  \\
&  &  &  &  &  &  &  &  \\
& \times & A(0,2) & \times & B(1,1) & \times & A(2,0) & \times &  \\
&  &  &  &  &  &  &  &  \\
& C(-1,2) & \times & D(0,1) & \times & C(1,0) & \times & D(2,-1) &  \\
&  &  &  &  &  &  &  &  \\
& \times & B(-1,1) & \times & A(0,0) & \times & B(1,-1) & \times &  \\
&  &  &  &  &  &  &  &  \\
\cdots & D(-2,1) & \times & C(-1,0) & \times & D(0,-1) & \times & C(1,-2) &
\cdots \\
\swarrow & \vdots &  &  &  &  &  & \vdots & \searrow
\end{array}
\end{equation}

Each cell on the worldsheet is labelled by the values of $(\sigma^{+},
\sigma^{-})$ at the center of the cell, divided by a factor of $\sqrt{2}$.
For example at the center of the cell labelled by $(1,2)$ the worldsheet
coordinates are $\sigma ^{+}=\sqrt{2},\ \ \ \sigma ^{-}=2\sqrt{2}$.
Furthermore the values of $(\tau ,\sigma )$ at any point on the sheet are
$\tau =(\sigma ^{+}+\sigma ^{-})/\sqrt{2}$ and $\sigma =(\sigma ^{+}-\sigma
^{-})/\sqrt{2}$. So, at the center of the cell labelled as $(m,n)$ the
worldsheet coordinates are $\tau =m+n,\quad \sigma =m-n$. The points inside
the cell $(m,n)$ are parametrized by $\sigma ^{\pm }$ in the ranges
\begin{equation}
\label{means}(m-{\frac 12})\sqrt{2}<\sigma ^{+}<(m+{\frac 12})\sqrt{2},
\qquad (n-{\frac 12})\sqrt{2}<\sigma ^{-}<(n+{\frac 12})\sqrt{2}.
\end{equation}

We define the periodic functions $\sigma _{per}^{\pm }$ that take the
following values in cell $(m,n)$
\begin{equation}
\label{period}
\begin{array}{l}
\sigma _{per}^{+}=\sigma ^{+}-m
\sqrt{2},\qquad \sigma _{per}^{-}=\sigma ^{-}-n\sqrt{2} \\
-\frac 1{\sqrt{2}}<\sigma _{per}^{\pm }<\frac 1{\sqrt{2}},
\qquad {\rm {periodic.}}
\end{array}
\end{equation}
This is the periodic saw-tooth function that has slope $+1.$

With these definitions, the solution (\ref{oldee}-\ref{fourflat})
 can be rewritten in the
following form:
\begin{equation}
\label{EQN}
\begin{array}{lll}
&  &  \\
A(0,0): & u_{00}=u_0+p^{+}\sigma _{per}^{+}, & v_{00}=v_0+p^{-}\sigma
_{per}^{-} \\
B(1,-1): & u_{1,-1}=u_0+p^{+}\sigma _{per}^{-}, & v_{1,-1}=v_0+p^{-}\sigma
_{per}^{+} \\
D(0,1): & u_{01}=(u_0+{\frac{p^{+}}{\sqrt{2}})}+p^{+}(\sigma
_{per}^{+}+\sigma _{per}^{-}), & v_{01}=v_0+
{\frac{p^{-}}{\sqrt{2}}} \\ C(1,0): & u_{10}=u_0+{\frac{p^{+}}{\sqrt{2}}}, &
v_{10}=(v_0+
{\frac{p^{-}}{\sqrt{2}})+}p^{-}(\sigma _{per}^{+}+\sigma _{per}^{-}) \\
B(1,1): & u_{11}=(u_0+2{\frac{p^{+}}{\sqrt{2}}})+p^{+}\sigma _{per}^{-}, &
v_{11}=(v_0+2
{\frac{p^{-}}{\sqrt{2}}})+p^{-}\sigma _{per}^{+} \\
A(2,0): & u_{20}=(u_0+2{
\ \frac{p^{+}}{\sqrt{2}}})+p^{+}\sigma _{per}^{+}, & v_{20}=(v_0+2
{\frac{p^{-}}{\sqrt{2}}})+p^{-}\sigma _{per}^{-} \\
D(2,1): & u_{21}=(u_0+3{\ \frac{p^{+}}{\sqrt{2}})}+p^{+}(\sigma _{per}^{+}+
\sigma _{per}^{-}), &
v_{21}=v_0+3
{\frac{p^{-}}{\sqrt{2}}} \\ C(1,2): & u_{12}=u_0+3{\frac{p^{+}}{\sqrt{2}}},
& v_{12}=(v_0+3
{\frac{p^{-}}{\sqrt{2}})}+p^{-}(\sigma _{per}^{+}+\sigma _{per}^{-}) \\
A(2,2): & u_{22}=(u_0+4{\frac{p^{+}}{\sqrt{2}}})+p^{+}\sigma _{per}^{+}, &
v_{22}=(v_0+4 {\frac{p^{-}}{\sqrt{2}}})+p^{-}\sigma _{per}^{-} \\
B(3,1):\, & u_{31}=(u_0+4{\frac{p^{+}}{\sqrt{2}}})+p^{+}\sigma _{per}^{-}, &
v_{31}=(v_0+4{\frac{p^{-}}{\sqrt{2}}})+p^{-}\sigma _{per}^{+}
\end{array}
\end{equation}

This list is the same solution as (\ref{old}-\ref{fourflat})
and it illustrates the
procedure for building the solution. Namely, first assign the patterns to
insure forward propagation, then impose periodicity at a fixed value of
$\tau ,$ e.g at $\tau =0,$ and then obtain the propagation into the future by
simply matching boundary conditions while increasing $\tau$.

The general solution in (\ref{olde}), that has more folds on the string,
defines a corresponding pattern of $A,B,C,D$ on the worldsheet. Given
correct boundary conditions in the asymptotic region, these patterns are
guarantied to correspond to strings that propagate forward
in time. This is understood intuitively by considering that
curved spacetime is a continuous deformation of flat spacetime except
for singularities, and that the flat solution is manifestly periodic
and forward propagating.
These decompositions of the worldsheet into cells is the key for
building the physical string solutions in curved spacetime. We emphasize
that we must consider precisely these patterns if we want to describe
the motion of strings prepared by observers in the asymptotic region.

\section{Boundary matching}
\setcounter{equation}{0}
\subsection{General approach for any metric}

We have seen that the four types of solutions are given in the cells defined
naturally by the periodic saw-tooth functions $\sigma _{per}^{\pm }$ :
\begin{equation}
\label{four}
\begin{array}{ll}
A:\qquad u=u(\sigma _{per}^{+}), & \quad v=v(\sigma _{per}^{-}) \\
B:\qquad u=\bar u(\sigma _{per}^{-}), & \quad v=\bar v(\sigma _{per}^{+}) \\
C:\qquad u=c_1, & \quad v=v(\bar a(\sigma _{per}^{+}),\bar b(\sigma
_{per}^{-})) \\
D:\qquad u=u(a(\sigma _{per}^{-}),b(\sigma _{per}^{+})), & \quad v=c_2.
\end{array}
\end{equation}
For a classical description of a string the time coordinate $T(\tau ,\sigma
)=(u+v)/\sqrt{2}$ must {\it increase} as a function of $\tau $ {\it for all}
$\sigma$. To achieve this we must start from a cell decomposition of the
worldsheet such that in each cell one of the solutions is valid, and then
connect them by continuity accross the boundaries. The patching of the
solutions must respect two important requirements:
\begin{equation}
\label{require}(i)\,\,periodicity\,\,\,in\,\,\,\sigma ,\quad and\quad
(ii)\,\,forward\,\,\,propagation.
\end{equation}
Satisfying these two requirements turns out to be rather non-trivial.
However, the folded string solutions of \cite{bbhp} provide a guide to
construct the curved spacetime solutions.
Without this guide it seems bewildering what solution corresponds to
each cell. As an example we start with the simplest yo-yo solution
(\ref{oldee}) in the asymptotic region, and use it to assign the
$A,B,C,D$ solutions to various patches according to (\ref{sol}).

The $A$ and $B$ solutions are present for any metric in target spacetime,
while the form of the $v(a,b)$ or $u(\bar a,\bar b)$ in the $C,\,D$
solutions depend on the metric $G(u,v)$ in target spacetime. Since there
still is the freedom to fix the remaining conformal gauge invariance, we fix
it by concentrating on solutions of type $A$ and $B$ that are independent of
the target spacetime metric. It is convenient to fix the gauge so as to
reproduce the well known solutions in flat spacetime in the asymptotic
region where the metric is flat anyway. Therefore for solutions of type $A$
and $B$ we take the gauge fixed form
\begin{equation}
\label{gauge}
\begin{array}{c}
A:\qquad u=p_A^{+}\sigma _{per}^{+}+q_A^{+},\quad v=p_A^{-}\sigma
_{per}^{-}+q_A^{-} \\
B:\qquad u=p_B^{+}\sigma _{per}^{-}+q_B^{+},\quad v=p_B^{-}\sigma
_{per}^{+}+q_B^{-}.
\end{array}
\end{equation}
The constants depend on the cell, and they can be chosen as initial
conditions on only one cell. It turns out that there is no more remaining
gauge degrees of freedom, and the functions $a,b,\bar a,\bar b$ in solutions
$C,D$ get fixed completely by the forward propagation in $\tau $, and take a
form which depends on the spacetime metric.

We start by assigning initial conditions in cell (0,0) by taking the $A$
type solution
\begin{equation}
\label{zerozero}u_{00}=u_0+p^{+}\sigma _{per}^{+},\qquad
v_{00}=v_0+p^{-}\sigma _{per}^{-}
\end{equation}
Then in the neighboring cell, by continuity in $\sigma ,$ the $B(1,-1)$
solution must be
\begin{equation}
\label{neighbors}u_{1,-1}=u_0+p^{+}\sigma _{per}^{-},\qquad
v_{1,-1}=v_0+p^{-}\sigma _{per}^{+}\,\,.
\end{equation}
This is sufficient to fix all the even and odd cells horizontally at $\tau
=0 $ because of periodicity in $\sigma $
\begin{equation}
\label{tauzero}
\begin{array}{c}
u_{2k,-2k}=u_0+p^{+}\sigma _{per}^{+},\qquad v_{2k,-2k}=v_0+p^{-}\sigma
_{per}^{-} \\
\\
u_{2k+1,-2k-1}=u_0+p^{+}\sigma _{per}^{-},\qquad
v_{2k+1,-2k-1}=v_0+p^{-}\sigma _{per}^{+}. \\
\end{array}
\end{equation}

Next we go one level up to cells $(1,0)$ and $(0,1)$. Consider the boundary
between (0,0) and (1,0) which is at $\sigma^+_{per}={\frac{1}{\sqrt{2}}},\ \
\sigma^-_{per} =any$. Matching the solutions of type $A$ and $C$ we find
\begin{equation}
\label{matchAC}u_{10}=u_0+{{\frac{p^{+}}{\sqrt{2}}}},\qquad v_{10}(a({\frac
1{\sqrt{2}}}),b(\sigma _{per}^{-}))=v_0+p^{-}\sigma _{per}^{-}.
\end{equation}
A similar matching of solutions of type $B$ and $C$ at the boundary between
(1,-1) and (1,0) gives
\begin{equation}
\label{matchBC}u_{10}=u_0+{{\frac{p^{+}}{\sqrt{2}}}},\qquad v_{10}(\bar
a(\sigma _{per}^{+}),\bar b(-{\frac 1{\sqrt{2}}}))=v_0+p^{-}\sigma
_{per}^{+}.
\end{equation}
Taken together these two conditions completely fix the functions $%
a_{10}(\sigma _{per}^{+})$ and $b_{10}(\sigma _{per}^{-})$, once the
function $v(a,b)$ is given for some target spacetime metric. Replacing these
functions back into $v(a,b)$ one completely fixes $v_{10}$. To make further
progress we must specify the metric and the corresponding solutions $u(\bar
a,\bar b),\,v(a,b).$

\subsection{2D black hole metric}

So, for the $SL(2,R)/R$ two dimensional black hole metric we have the form
$v_{10}=(1-a_{10}b_{10})/u_{10}$. Using the above procedure we find
\begin{equation}\label{vonezero}
\begin{array}{rl}
u_{10} & = u_0 + \frac{p^+}{\sqrt{2}} \\
v_{10} & = \frac{1}{u_0+\frac{p^+}{\sqrt{2}}} \left\{ 1-\frac{[1-
   (u_0+\frac{p^{+}}{\sqrt{2}})(v_0+p^{-}\sigma _{per}^{+})]\times
   [1-(u_0+\frac{p^{+}}{\sqrt{2}})(v_0+p^{-}\sigma _{per}^{-})]}
   {[1-(u_0+\frac{p^{+}}{\sqrt{2}})(v_0-\frac{p^{-}}{\sqrt{2}})]}\right\}
\end{array}
\end{equation}
A similar matching procedure for $B(-1,1)\nearrow D(0,1)\nwarrow A(0,0)$
gives
\begin{equation}
\label{uvzeroone}
\begin{array}{rl}
v_{01} & =v_0 + {{\frac{p^{-}}{\sqrt{2}}}} \\
u_{01}(\sigma_{per}^+,\sigma_{per}^-) & = \frac{1}{v_0+\frac{p^-}{\sqrt{2}}}
   \left\{ 1-{\frac{[1-(v_0+{{\frac{p^{-}}{\sqrt{2}}}})
   (u_0+p^{+}\sigma _{per}^{+})]\times [1-(v_0+{\frac{p^{-}}{\sqrt{2}}})
   (u_0+p^{+}\sigma _{per}^{-})]}{[1-(v_0+{\frac{p^{-}}{\sqrt{2}}})
   (u_0-{\frac{p^{+}}{\sqrt{2}}})]}}\right\}
\end{array}
\end{equation}
By periodicity these solutions are extended to the rest of the even and odd
cells horizontally at the same value of $\tau $. So, we may write
\begin{equation}
\label{tauone}
\begin{array}{ll}
u_{2k+1,-2k}=u_0+{\frac{p^{+}}{\sqrt{2}}}, & \qquad
v_{2k+1,-2k}=v_{10}(\sigma _{per}^{+},\sigma _{per}^{-}) \\
u_{-2k,2k+1}=u_{01}(\sigma _{per}^{+},\sigma _{per}^{-}), & \qquad
v_{-2k,2k+1}=v_0+\frac{p^{-}}{\sqrt{2}}
\end{array}
\end{equation}

We next climb one more level in $\tau $ and consider the cells $B(1,1),\
A(2,0)$. Matching boundaries $D(0,1)\nearrow B(1,1)\nwarrow C(1,0)$ and $%
C(1,0)\nearrow A(2,0)\nwarrow D(2,-1)$ we get the solution
\begin{equation}
\label{tautwo}
\begin{array}{l}
u_{11}(\sigma _{per}^{-})=
{\frac 1{v_0+{{\frac{p^{-}}{\sqrt{2}}}}}}\left\{ 1-{\frac{[1-(v_0+{{\frac{%
p^{-}}{\sqrt{2}}}})(u_0+{{\frac{p^{+}}{\sqrt{2}}}})]}{[1-(v_0+{{\frac{p^{-}}{%
\sqrt{2}}}})(u_0-{{\frac{p^{+}}{\sqrt{2}}}})]}}[1-(v_0+{\frac{p^{-}}2}%
)(u_0+p^{+}\sigma _{per}^{-})]\right\} \\  \\
v_{11}(\sigma _{per}^{+})=
{\rm {similar\ to\ }u_{11}\ {but\ replace\ }\ \ u_0\leftrightarrow v_0\ ,\ \
p^{+}\leftrightarrow p^{-},\ \ \sigma _{per}^{-}\rightarrow \sigma
_{per}^{+}\ ,} \\ u_{20}(\sigma _{per}^{+})=
{\rm {same\ as}\ u_{11}\ \ {but\ replace}\ \sigma _{per}^{-}\rightarrow
\sigma _{per}^{+}} \\ v_{20}(\sigma _{per}^{-})={\rm {same\ as}\ v_{11}\ \ {%
but\ replace}\ \ \sigma _{per}^{+}\rightarrow \sigma _{per}^{-}}
\end{array}
\end{equation}
By periodicity in $\sigma $ this solution is extended horizontally to all
cells at the same $\tau $. Thus
\begin{equation}
\label{tautwoo}
\begin{array}{ll}
u_{2k+1,-2k+1}=u_{11}(\sigma _{per}^{-}), & \qquad
v_{2k+1,-2k+1}=v_{11}(\sigma _{per}^{+}), \\
u_{2k+2,-2k}=u_{20}=u_{11}(\sigma _{per}^{+}), & \quad \quad
v_{2k+2,-2k}=v_{20}=v_{11}(\sigma _{per}^{-}),
\end{array}
\end{equation}

In the next step we match $A(0,2)\nearrow C(1,2)\nwarrow B(1,1)$ and $%
B(1,1)\nearrow D(2,1)\nwarrow A(2,0)$. The result must be the same as
starting in the cell $A(2,2)$ and matching boundaries backward in $\tau $
with cell $C(1,2).$ Therefore the result in $C(1,2)$ can be written in the
form that relates to the constants in cell $A(2,2),$ that is
\begin{equation}
\label{tauthree}
\begin{array}{cl}
\quad \quad \quad \quad \quad u_{12} & =\tilde u_0-
\frac{\tilde p^{+}}{\sqrt{2}} \\ v_{12}(\sigma _{per}^{+},\sigma _{per}^{-})
& =
{\frac 1{\tilde u_0-{{\frac{\tilde p^{+}}{\sqrt{2}}}}}}\left\{ 1-{\frac{%
[1-(\tilde u_0-{{\frac{\tilde p^{+}}{\sqrt{2}}}})(\tilde v_0+\tilde
p^{-}\sigma _{per}^{+})]\times [1-(\tilde u_0-{{\frac{\tilde p^{+}}{\sqrt{2}}%
}})(\tilde v_0+\tilde p^{-}\sigma _{per}^{-})]}{[1-(\tilde u_0-{{\frac{%
\tilde p^{+}}{\sqrt{2}}}})(\tilde v_0+{{\frac{\tilde p^{-}}{\sqrt{2}}}})]}}%
\right\} \\ \quad \quad \quad \quad \quad v_{21} & =\tilde v_0-
{{\frac{\tilde p^{-}}{\sqrt{2}}}} \\ u_{21}(\sigma _{per}^{+},\sigma
_{per}^{-}) & ={\rm {similar\ to}\ v_{12}\ \ {but\ substitute}\ \tilde
u_0\leftrightarrow \tilde v_0,\ \ \tilde p^{+}\leftrightarrow \tilde p^{-}}
\end{array}
\end{equation}
where we have defined $A(2,2)$ by $u_{22}=\tilde u_0+\tilde p^{+}\sigma
_{per}^{+}$ and $v_{22}=\tilde v_0+\tilde p^{-}\sigma _{per}^{-}.$ By
comparing this with the results of $A(0,2)\nearrow C(1,2)\nwarrow B(1,1)$
and $B(1,1)\nearrow D(2,1)\nwarrow A(2,0)$, the constants $\tilde u_0,\
\tilde v_0,\ \tilde p^{+},\ \tilde p^{-}$ are related to the original
initial conditions at $\tau =0$:
\begin{equation}
\label{uvprime}
\begin{array}{l}
\tilde u_0-
{\frac{\tilde p^{+}}{\sqrt{2}}}={\dfrac 1{v_0+{{\frac{p^{-}}{\sqrt{2}}}}}}%
\left\{ 1-{\dfrac{[1-(u_0+{{\frac{p^{+}}{\sqrt{2}}}})(v_0+{{\frac{p^{-}}{%
\sqrt{2}}}})]^2}{[1-(u_0-{{\frac{p^{+}}{\sqrt{2}}}})(v_0+{{\frac{p^{-}}{%
\sqrt{2}}}})]}}\right\} \\  \\
\tilde v_0-
{\frac{\tilde p^{-}}{\sqrt{2}}}={\rm {same\ form\ but\ interchange}\
u_0\leftrightarrow v_0,\ \ \ p^{+}\leftrightarrow p^{-}} \\  \\
\tilde p^{+}=\left\{
\dfrac{[(1-(u_0+\frac{p^{+}}{\sqrt{2}})(v_0+{{\frac{p^{-}}{\sqrt{2}}}}%
)]^2-2p^{+}p^{-}}{[1-(u_0-{\frac{p^{+}}{\sqrt{2}}})(v_0+{\frac{p^{-}}{\sqrt{2%
}}})]^2}\right\} \,\,\,p^{+} \\  \\
\tilde p^{-}=
{\rm {same\,\,form\ but\ interchange}\ \ u_0\leftrightarrow v_0,\
p^{+}\leftrightarrow p^{-}} \\
\end{array}
\end{equation}
The result in cell $A(2,2)$ has the same form as the cell $A(0,0)$ given in (%
\ref{zerozero}) except for replacing $\tilde u_0,\tilde v_0,\tilde
p^{+},\tilde p^{-}$ instead of $u_0,v_0,p^{+},p^{-}$ . Therefore the
procedure can now be repeated by using the constants $\tilde u_0,\ \tilde
v_0,\ \tilde p^{+},\ \tilde p^{-}$as new initial conditions. So, the forward
propagation of the solution to future times $\tau \rightarrow \tau +4$ is
done through the formulas in (\ref{uvprime}) which serve the role of a {\it {%
transfer operation }}analogous to the transfer matrix of a lattice theory%
{\it . }

In the asymptotic region $u_0\rightarrow \infty ,\ v_0\rightarrow -\infty $
(where $G\to $ flat) the formulas reduce to the results for a flat metric.
Namely,
\begin{equation}
\label{flat}\tilde p^{\pm }=p^{\pm },\qquad \tilde u_0=u_0+4{\frac{p^{+}}{%
\sqrt{2}}},\qquad \tilde v_0=v_0+4{\frac{p^{-}}{\sqrt{2}}}
\end{equation}
and the propagation in flat space coincides with (\ref{EQN}) as given in
the previous section.

\subsection{Transfer operator and invariant minimal area}

Although we have derived this result by choosing a gauge in eq. (\ref{gauge}%
) it is easy to verify that the constants $u_{10},v_{01},u_{12},v_{21}$ in
the $C,D$ type cells are independent of the gauge. That is, even if we had
started from arbitrary functions in cell $A(0,0)$ , i.e. $u_{00}=u(\sigma
_{per}^{+})$ and $v_{00}=v(\sigma _{per}^{-}),$ the values of these
functions at the boundaries of the cell is what defines $u_0\pm p^{+}/\sqrt{2%
},\,\,v_0\pm p^{-}/\sqrt{2}.$ Then the values of the constants
\begin{equation}
u_{\pm 1,0}=u_0\pm p^{+}/\sqrt{2},\quad v_{0,\pm 1}=v_0\pm p^{-}/\sqrt{2}
\end{equation}
have identical expressions, independent of the gauge. As we will see below,
these boundary constants and their transfer to future times via the transfer
operation completely determine the gauge invariant motion of the folds (or
end points for an open string). The physical motion of the remainder of the
string in between the folds is infered from the motion of the folds. Since
we can choose $u(\sigma _{per}^{+})$ and $v(\sigma _{per}^{-})$ arbitrarily,
it is evident that the motion of the intermediate points is gauge dependent,
and therefore has no physical meaning. The physically significant motion is
the motion of the folds (or end points). These issues were understood for
the flat case by BBHP.

In order to give a convenient expression for the transfer operation it is
more convenient to concentrate on the gauge invariant constants in the $C,D$
type patches. The $C$ patches with constant values of $u$ are isomorphic to
the cells $(2k+1,0)$ by periodicity at $\tau =2k+1$, and similarly the $D$
patches with constant values of $v$ are isomorphic to the cells $(0,2k+1).$
Therefore we may write%
\begin{equation}
u_{2k+1}\equiv u_{2k+1+2l,-2l}\quad ,\quad v_{2k+1}\equiv
v_{-2l,2k+1+2l}\quad ,\quad k,l=0,\pm 1,\pm 2,\pm 3,\cdots
\end{equation}
The four constants $u_{\pm 1},v_{\pm 1}$ for $k=-1,0$ and $l=0$ are
equivalent to the four constants that define the initial conditions $%
u_0,v_0,p^{\pm }.$ The transfer operation given in (\ref{uvprime}) relate
the initial values ($u_{\pm 1}$, $v_{\pm 1})$ to the future values ($%
u_3,v_3),$ at the later time $\tau \rightarrow \tau +2$, etc. It is not
difficult to rewrite the transfer operation as a recursion relation for any $%
k.$ Thus, given four constants ($u_{2k-1}$, $v_{2k-1})$ and ($u_{2k+1}$, $%
v_{2k+1})$ at some $k$ we find the action of the transfer operation by
deriving the future constants for $k\rightarrow k+1,$ i.e. ($u_{2k+3}$, $%
v_{2k+3}):$
$$
(u_{2k+3}-u_{2k+1})=
\frac{(1-u_{2k+1}v_{2k+1})}{(1-u_{2k-1}v_{2k+1})}(u_{2k+1}-u_{2k-1})
$$
\begin{equation}
\label{transfer}
(v_{2k+3}-v_{2k+1})=\frac{(1-u_{2k+1}v_{2k+1})}{(1-u_{2k+1}v_{2k-1})}
(v_{2k+1}-v_{2k-1})
\end{equation}
This provides a recursion relation that gives the progress of the folds in
discrete steps. We will call this map the {\it transfer matrix.}

The motion of fold-1 at $\sigma =-1$ is obtained as follows. At $\tau =0$
fold-1 is identified as the cross ($\times )$ between cells $C(-1,0)$ and $%
D(0,1).$ From the constants in those cells we infer that the spacetime
location of fold-1 is $(u_{-1,0}=u_{-1},\,\,\,\,v_{01}=v_1).$ At $\tau =2$
fold-1 is identified as the cross between $D(0,1)$ and $C(1,2).$ Therefore,
the spacetime location has moved to $(u_{12}=u_3\,\,\,,v_{01}=v_1).$ At $%
\tau =4$ it is at the cross between $C(1,2)$ and $D(2,3)$. Therefore it has
moved to the spacetime point $(u_{12}=u_{3\,\,\,},v_{23}=v_5),$ and so on.
In a similar way we can find the spacetime location of fold-2 at $\sigma =1$%
, as well as the location of the midpoint between the two folds at $\sigma
=0,2$ that represents the center of mass of the string. The successive
spacetime locations of these points are as follows
\begin{equation}
\label{successive}
\begin{array}{lll}
\text{fold-1}\quad(\sigma =-1) & \text{midpoint}\quad(\sigma =0,2)
   & \text{fold-2 \quad }(\sigma =1) \\
(u_{-1,0}=u_{-1}\,,v_{01}=v_1)_{\tau=0} & (u_{10}=u_1,\,v_{01}=v_1)_{\tau =1}
   & (u_{10}=u_1,\,v_{0,-1}=v_{-1})_{\tau =0} \\
(u_{12}=u_3,\,\,v_{01}=v_1)_{\tau =2} & (u_{12}=u_3,\,v_{21}=v_3)_{\tau =3}
   & (u_{10}=u_1,\,\,v_{21}=v_3)_{\tau =2} \\
(u_{12}=u_3,\,v_{23}=v_5)_{\tau =4} & (u_{32}=u_5,\,\,v_{23}=v_5)_{\tau =5}
   & (u_{32}=u_5,\,v_{21}=v_3)_{\tau=4} \\
\vdots & \vdots & \vdots
\end{array}
\end{equation}
{}From these values we can construct one full oscillation of the string. Note
that we have the freedom to choose four initial constants. The four initial
constants already determine half of the motion up to $\tau =2$. We then
iterate according to (\ref{transfer}) to compute the future motion. The
resulting motion of the string is plotted in the figures with the help of a
computer. We will describe it in section 6.

We now make an important observation. If we consider the rectangular shapes
traced by the minimal surfaces, we may compute approximately the invariant
area of rectangle $k$ as
\begin{equation}
\label{area}A_k=\frac{(\Delta u)_k\,\,(\Delta v)_k}{(1-u_0v_0)_k}
\end{equation}
where ($\Delta u)_k,\,(\Delta v)_k$ are the sides of the squares and ($%
u_0,v_0)_k$ are the values of $u,v$ in the middle of rectangle $k$. We will
call this quantity the {\it lattice minimal area} to distinguish it from the
actual minimal area $\int d\tau \,d\sigma $ ($\partial _{+}u\partial
_{-}v+\partial _{-}u\partial _{+}v)/(1-uv)$. It is evident that the lattice
minimal area is an approximation to the actual minimal area. The remarkable
observation is that the lattice minimal area $A_k$ as defined in (\ref{area}%
) is actually an invariant of the motion. That is we can show that
\begin{equation}
\label{invariant}A=4\frac{(u_{2k+1}-u_{2k-1})(v_{2k+1}-v_{2k-1})}{%
[4-(u_{2k+1}+u_{2k-1})(v_{2k+1}+v_{2k-1})]}
\end{equation}
is an invariant
under the action of the transfer matrix.
To prove it we only need to consider two neigboring values
of $k$ and compare $A_k$ to $A_{k+1}$ that are related to each other by the
transfer matrix. By using the formulas in (\ref{transfer}) it takes a little
algebra to verify our assertion.

The lattice minimal area as well as the transfer matrix are reparametrization
invariants (i.e. independent of the conformal gauge) since they are functions
of only the positions of the folds (or end points).

\subsection{Generalizations}

The general flat spacetime solution generates a more complicated periodic
$A,B,C$, $D$ pattern than (\ref{sol}), and that is the pattern that must be
used for the general curved spacetime solution that connects to the flat
spacetime motion as an initial condition. The role of the $A,B,C,D$ pattern
is to insure that the global time coordinate $T=(u+v)/\sqrt{2}$ is an
increasing function of $\tau $ for all $\sigma .$ The pattern is defined by
the flat solution written in the form
\begin{equation}
\label{general}
\begin{array}{c}
u(\sigma ^{+},\sigma ^{-})=u_0+
\frac{p^{+}}2\left[ (\sigma ^{+}+f(\sigma ^{+}))+(\sigma ^{-}-g(\sigma
^{-}))\right] \\ v(\sigma ^{+},\sigma ^{-})=v_0+\frac{p^{-}}2\left[ (\sigma
^{+}-f(\sigma ^{+}))+(\sigma ^{-}+g(\sigma ^{-}))\right]
\end{array}
\end{equation}
where $f(\sigma ^{+})$ and $g(\sigma ^{-})$ are any {\it periodic functions}
with slopes $f^{\prime }(\sigma ^{+})=\pm 1$ and $g^{\prime }(\sigma
^{-})=\pm 1.$ The slope can change discontinuously any number of times and
at arbitrary locations $\sigma^+_i,\sigma^-_j$ within the basic intervals $%
-\frac 1{\sqrt{2}}\leq \sigma ^{\pm }\leq \frac 1{\sqrt{2}}.$ For example we
can define the analog of $\sin (k\sigma ^{\pm })$, as $s_k(\sigma
^{+})=(-1)^m\sigma _{per}^{\pm }$, and $s_{\tilde k}(\sigma
^{-})=(-1)^m\sigma _{per}^{-}$ with periods of ($\sqrt{2}/k,\sqrt{2}/\tilde
k)$ for ($\sigma _{per}^{+},\sigma _{per}^{-})$ respectively. By the
arguments of BBHP one sees that the choice $f(\sigma ^{+})=s_k(\sigma ^{+})$
and $g(\sigma ^{-})=s_{\tilde k}(\sigma ^{-})$ gives the flat spacetime
solution that corresponds to {\it normal modes} for folded strings. This
choice of functions describes strings that are folded in equal lengths, $2k$
times for left movers and $2\tilde k$ times for right movers. One could
discuss the normal modes or any arbitrary motion.

The curved spacetime solution based on the $A,B,C,D$ patterns generated by
the form (\ref{general}) will have this expression as the approximate
initial condition in the asymptotically flat region. Therefore, the
procedure for obtaining the curved spacetime solution is to first build the
worldsheet lattice defined by the functions $f(\sigma ^{+})$ and $g(\sigma
^{-}).$ This means drawing the $45^o$ lines along $\sigma ^{\pm }$ according
to the patterns defined by these functions. Next rewrite the solution (\ref
{general}) in the forms $A,B,C,D$ as in the example (\ref{EQN})
and figure out the $A,B,C,D$ pattern on the lattice. Then, instead of the
flat $A,B,C,D$ functions substitute the functions (\ref{fourr}) for the
appropriate curved spacetime metric and demand periodicity in $\sigma
\rightarrow \sigma +4$ at $\tau =0.$ This defines the initial condition at $%
\tau =0$ and the expected patterns for all future values of $\tau .$ Then by
matching the solutions at the boundaries one fixes the remaining boundary
parameters. It is best to rewrite these parameters as the $(u,v)$ locations
of the folds at successive switching points as described in the discussion
before (\ref{transfer}). By carrying out this procedure for a complete
periodic motion of all the folds one obtains the transfer operation.

These arguments show that a complete classification of all the {\it physical}
classical solutions is obtained by the complete classification of all the
boundary conditions in the asymptotically flat region, and those are given
by (\ref{general}). Each classical solution for a given metric defines a
transfer matrix.

Furthermore, it is evident that for each distinct spacetime metric $G_{\mu
\nu }(x)$ we obtain a different transfer matrix. This introduces an
interesting idea that connects geometry on the one hand and and lattice-like
transfer matrices for minimal surfaces on the other hand.

\section{String falling into the black hole}
\setcounter{equation}{0}

Using the solution obtained above we now describe what happens to a
relativistic string which is falling into the black hole. We will discuss
explicitely only the case of the closed string with two folds (or the open
string without folds), for which we have given detailed expressions in the
previous section. This is sufficient to understand the main physical
phenomena.

The two dimensional motion described here corresponds to a motion in 4
dimensions in the radial and time coordinates $(r,t).$ Thus, we may imagine
a string streched along the radial direction at fixed angles $(\theta ,\phi
) $ , and then falling into the black hole in the radial direction while
performing longitudinal oscillations. This may approximate an extended
object with internal degrees of freedom, such as a star or gas, that
oscillates under the influence of internal forces while falling into the
black hole.

1 -- Starting with a streched closed string in the asymptotic region, we can
now follow its development in time by using our classical solution in the 2D
black hole metric. The solution is plotted in target space $(u,v)$
coordinates in Figs.1,2,3. These three figures, taken together correspond to
the motion resulting from the same initial condition. Note the change in
scale from one figure to the next. The results are not what we would have
guessed {\it a priori}. At first the string begins to oscillate as in the
flat spacetime case, but the gravitational attraction of the black hole also
forces it to move towards it, as expected. The combined oscillatory and
dragging motion is almost as in eq.(\ref{oldee}) as long as the string
remains far away from the horizon. This produces a minimal surface in curved
spacetime similar to the one in flat spacetime. As the string passes the
horizon and approaches the black hole, it shrinks in average size, and the
{\it \ invariant lattice minimal area} that it sweeps in one oscillation
stays constant.

2 -- The string with generic initial conditions does not stop at the black
hole singularity, but rather, it continues its journey into a second sheet
of spacetime that is provided by the $SL(2,R)/R$ manifold. Once it passes to
the second sheet it begins to expand again and resumes its oscillatory
motion while {\it moving away from the singularity}. Thus, in the second
sheet the singularity is interpreted as a white hole. This motion continues
until the string reaches the other branch of the singularity to which it is
attracted. It shrinks in size while passing trough it and emerges back into
the first sheet (or a third sheet, depending on interpretation). Thus, the
second branch of the singularity, which is a black hole in the second sheet,
is a white hole in the first sheet (or third sheet). The motion continues,
and the string can come back to the region {\it outside the horizons} of
both branches of the singularity, where it started initially. If there are
only two sheets there is the possibility of making closed timelike curves.
However, in the quantum theory it is not necessary to interpret the
emergence from the white hole as coming back to the first sheet. Rather, it
would be interpreted as a third sheet if the wavefunction $\psi (u,v)$ does
not come back to the same value. In fact, in a quantum state that has an
irrational ``magnetic'' quantum number $m$ of $SL(2,R),$ the wavefunction $%
\psi _{jm}(u,v)$ never comes back to the same value\footnote{%
The wavefunction is simply the diagonal entries in the representation of the
group element $D_{m,m^{\prime }}^j(g)$ , with $m+m^{\prime }=0$ because of
gauge invariance \cite{ibsfexa}.}. For such states we have to interpret the
{\it covering manifold} as having an infinite number of sheets, which is
analogous to the Reissner-Nordstrom black hole manifold. The multisheet
structure is further clarified in item 6 below.

3 -- What happens precisely at the black hole? Does the string shrink to a
point before falling into the black hole, as it seems to be the case in
Figs.1,2? The answer is no, as seen in Fig.3. The string is smoothly
swallowed by the black hole like a spagetti, but the end inside reaches into
the forbidden region beyond the singularity! When the second end of the
string catches up as the string shrinks to a point inside the forbidden
region, the string snaps back and begins to expand into the second sheet of
the black hole. This is a strange behavior that point particles cannot
perform. Its origin must be related to the wave nature of the string. Like
wave phenomena in quantum mechanics, the wave motion of the string seems to
permit it to penetrate a barrier that cannot be reached by classical motion
of point particles. This is perhaps the most interesting surprize of our
analysis.

4 -- As seen in Fig.3 the four corners of the ingoing rectangle are
sufficient to infer the motion of the string as it is absorbed by the black
hole. Is it possible to eliminate the strange behavior by having the far
corner coincide with the black hole? This requires fixing two constants to
special values. Recall that the initial conditions for the yo-yo solution is
specified by just four numbers (the positions and velocities of the end
points). Therefore, generic initial conditions used by an observer far away
from the black hole will not satisfy the special requirements, and the
string will generically perform the strange behavior.

5 -- From the remarks in number 4 we can see that we have the ability of
choosing initial conditions to arrange for a sufficiently long string, such
that, while it is being swallowed by the black hole, one of its ends is
inside the black hole and its other end is outside the horizon. This raises
questions on whether one might be able to extract information about the
black hole by using a long string as a probe, instead of relying only on
light rays (which cannot escape the black hole).

6 -- One may also start with a string whose initial conditions are specified
in the ``bare singularity region''. We find that such a string {\it never}
hits the singularity. An example of its motion is given in Fig.4. Its
general average behavior is similar to the motion of the massive particle,
for which an analytic expression is given in (\ref{massivegeo}) below. As seen
from these expressions a massless particle can reach the singularity, but a
massive particle cannot do it. The string is a massive state, and this is the
explanation for not hitting the singularity. Of course, in the bare
singularity region that adjoins the white hole the story is different. In
that region the massive particle does fall into the singularity and then
moves into the second sheet. Similarly, a string follows a similar route, but
at the singularity, its minimal area tunnels into the physical region
outside of the white hole. This tunneling is similar to the one described in
items 3,4,5 above, but now occuring from the other side.

7 -- In order to understand the multi-sheet nature of the manifold, as well
as the general behaviour of the solution in the various black hole regions,
it is useful to investigate the motion of a string of zero size, or a
particle. Its equations of motion correspond to the geodesic equations of a
point particle, and for any conformal metric it is given by
\begin{equation}
\label{pointgeo}\partial_\tau (x^\mu G)=\frac 12\partial _\tau x\cdot
\partial_\tau x \frac{\partial G}{\partial x^\nu}\eta^{\mu \nu}
\end{equation}
This equation follows from the string equations by specializing to the
center of mass motion, by doing dimensional reduction (ignoring the $\sigma$
dependence). For the 2D black hole we use the lightcone coordinates $x^\mu
=(u,v)$ and $G(u,v)=(1-uv)^{-1}.$ These equations are solved by applying the
techniques of \cite{ibsfglobal} for $SL(2,R)/R$ which give the time
dependence of the $SL(2,R)$ group element for a point particle

\begin{equation}
\label{groupgeodesic}g(\tau) = \left(
\begin{array}{cc}
u(\tau) & a(\tau) \\
-b(\tau) & v(\tau)
\end{array}
\right)
\end{equation}
where
\begin{equation}
\label{massivegeo}
\begin{array}{c}
u(\tau) = e^{ \alpha \tau} \left\{ \,u_0\,\cosh (\gamma \tau)-[u_0\,\alpha
-a_0\,p^{+}]\,
\frac{1}{\gamma} \,\sinh (\gamma \tau)\right\} \\ v(\tau) = e^{-\alpha \tau
}\left\{ \,v_0\,\cosh (\gamma \tau)+[v_0\,\alpha +b_0\,p^{-}]\,
\frac{1}{\gamma} \,\sinh (\gamma \tau)\right\} \\ a(\tau) = e^{ \alpha \tau
}\left\{ \,a_0\,\cosh (\gamma \tau)+[a_0\,\alpha -u_0\,p^{-}]\,
\frac{1}{\gamma} \,\sinh (\gamma \tau)\right\} \\ b(\tau) = e^{-\alpha \tau
}\left\{ \,b_0\,\cosh (\gamma \tau)-[b_0\,\alpha +v_0\,p^{-}]\,
\frac{1}{\gamma} \,\sinh (\gamma \tau)\right\} \\ \smallskip \alpha \equiv
\frac 12 (\frac{u_0}{a_0}p^{-}-\frac{v_0}{b_0} p^{+}),\quad \gamma \equiv
[\alpha^2-\,p^{+}p^{-}]^{1/2},\quad \\ \smallskip ds^2/d\tau
^2=(1-uv)^{-1}\dot u\,\dot v=p^{+}p^{-}.
\end{array}
\end{equation}
The diagonal entries in $g(\tau)$ give the desired solutions to the geodesic
equations. This can be checked directly by substitution in the geodesic
equation (\ref{pointgeo}). The constants $u_0,v_0,p^{\pm}$ define the
initial conditions, and $u_0v_0+a_0b_0=1$ is the $SL(2,R)$ determinant
condition. When $\,p^{+}p^{-}$ is positive the geodesic represents the
motion of a massive particle. When either $p^{+}$ or $p^{-}$ is zero, it
corresponds to a massless particle that moves along light-like trajectories
(either $u$ or $v$ reduces to constant).

As seen from the explicit solution, for the massive particle
, in the $u_0v_0<1$ physical region, there are
initial conditions that correspond to an imaginary $\gamma$. In that case $%
u(\tau ),v(\tau )$ are oscillating functions, indicating backward
propagation in time for certain parts of the motion. This is understood as
moving from sheet to sheet in the $SL(2,R)/R$ manifold.

These pointlike solutions are useful to understand the general trend of the
motion of the string in all regions $uv<1$ or $uv>1$.
The overall string behaves like a massive state, and
therefore we should expect its motion to have some similarities to the
motion of massive particles, except for the wave-like phenomena that became
manifest in the vicinity of the black hole. Indeed the global trends of the
motion plotted in the figures are similar to the trends given by the
analytic expressions in (\ref{massivegeo}) for the massive particle case.
This is so despite the fact that the folds (or end points) on the string
move just like massless particles (always parallel to the $u$ or $v$ axes).
However, the folds are not free particles and they change direction abruply
under the influence of the string.

8 -- It is also possible to consider the sigma model directly without
relating it to the $SL(2,R)/R$ manifold. In this case one is not necessarily
commited to a given number of sheets. In particular one might want to insist
that the motion of Figs. 1,2,3 are all on the same sheet. Let us suppose the
string reaches the black hole when $\tau =\tau _0,$ i.e. $u(\tau _0,\sigma )$
$v(\tau _0,\sigma )=1.$ Then for $\tau >\tau _0$ the time coordinate $T(\tau
,\sigma )=(u+v)/\sqrt{2}$ {\it decreases }since there is only one sheet. To
interpret the backward motion physically we need to invoke an anti-string
moving forward in time on the same sheet, and annihilating the original
string in the vicinity of the black hole. Similarly pairs of strings are
emitted at the white hole. If this interpretation is adopted, it turns out
that there can be no solutions with single strings moving only forward in
time for all parts of the string in the $uv<1$ region. We find always
anti-strings together with strings in parts of the motion. This situation is
analogous to the Klein paradox in the interacting Klein-Gordon or Dirac
equations. As is well known in those cases, when strong fields are present
the classical equations produce phenomena that seem to violate physical
intuition. The physics behind these phenomena is the production of pairs of
particles and anti-particles due to the strong fields. When such strong
fields are present the Klein paradox is the signal that the correct physics
can no longer be fully described by the corresponding classical equations,
and that the proper formalism is quantum field theory. However, classical
solutions are still useful in obtaining at least qualitative physical
information.

9 -- We may repeat the analysis for the string motion in any spherically
symmetric curved spacetime in the the time-radial subspace at fixed $d\theta
=d\phi =0$. Interesting possibilities include cosmological spacetimes as
well as black hole type spacetimes. As discussed in section 2, the string
solutions have the same $A,B,C,D$ structure for every spacetime metric
in 2D. The dependence on the metric appears through the functions $%
v(a,b),\,\,u(a,b)$. In particular, as an example, one can now analyze the
classical string motions falling into the Schwarzschild black hole. We
expect general phenomena that are similar to the ones described in this
paper, but the details near the singularity may be somewhat different. If
one admits a second sheet (defined with negative $r$), the string goes into
the second sheet, otherwise with a single sheet there must be pairs of
strings and anti-strings annihilated at the black hole, as argued in item 8.

10 -- Is there message from all this for the information paradox in black
holes? Evidently, we seem to be learning something that was not suspected by
studying only point particles. Given these string phenomena, the discussions
about the information paradox in black hole physics may also need revision.

\newpage

\subsection{\bf Figure captions}

Fig.\thinspace 1. String approaching the black hole. The invariant minimal
area per period is a constant of motion.

Fig.\thinspace 2. The ingoing string on the first sheet meets the black hole
and moves out to the second sheet.

Fig.\thinspace 3. The string minimal area tunnels to the forbidden region
beyond the black hole.

Fig.\thinspace 4. The string in the dual region (naked singularity region)
does not hit the singularity.

\newpage

\putfig{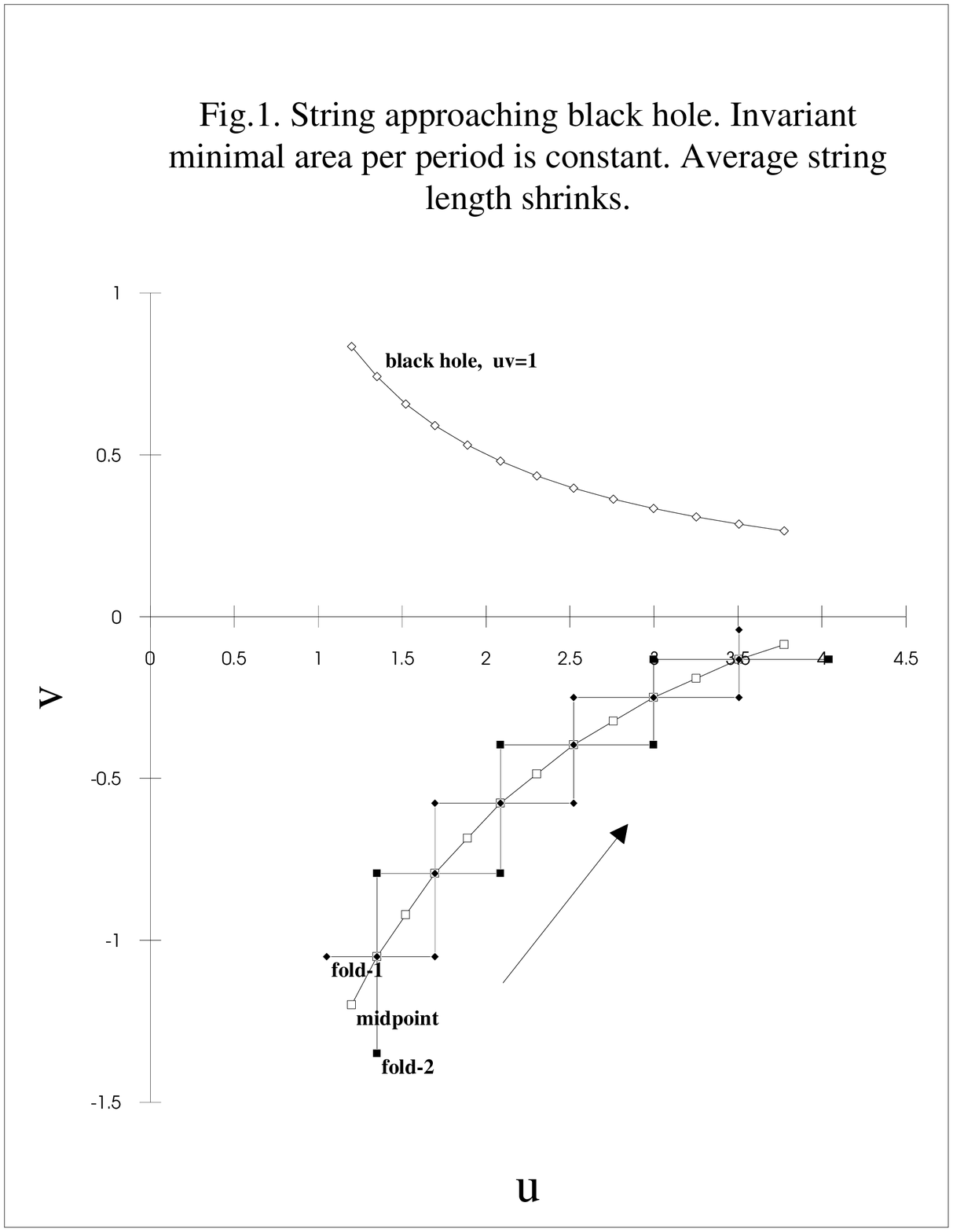}{}{150mm}

\putfig{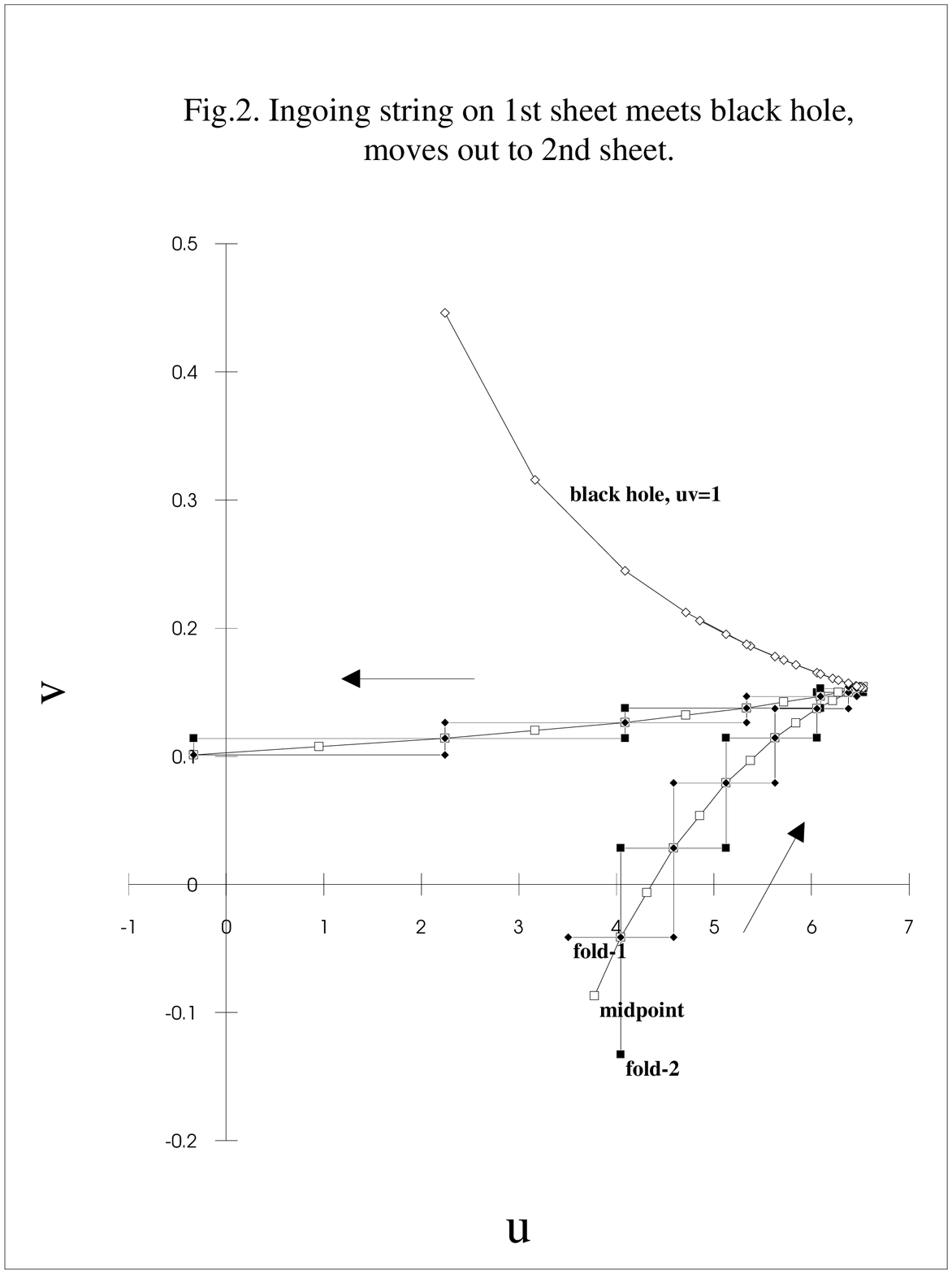}{}{150mm}

\putfig{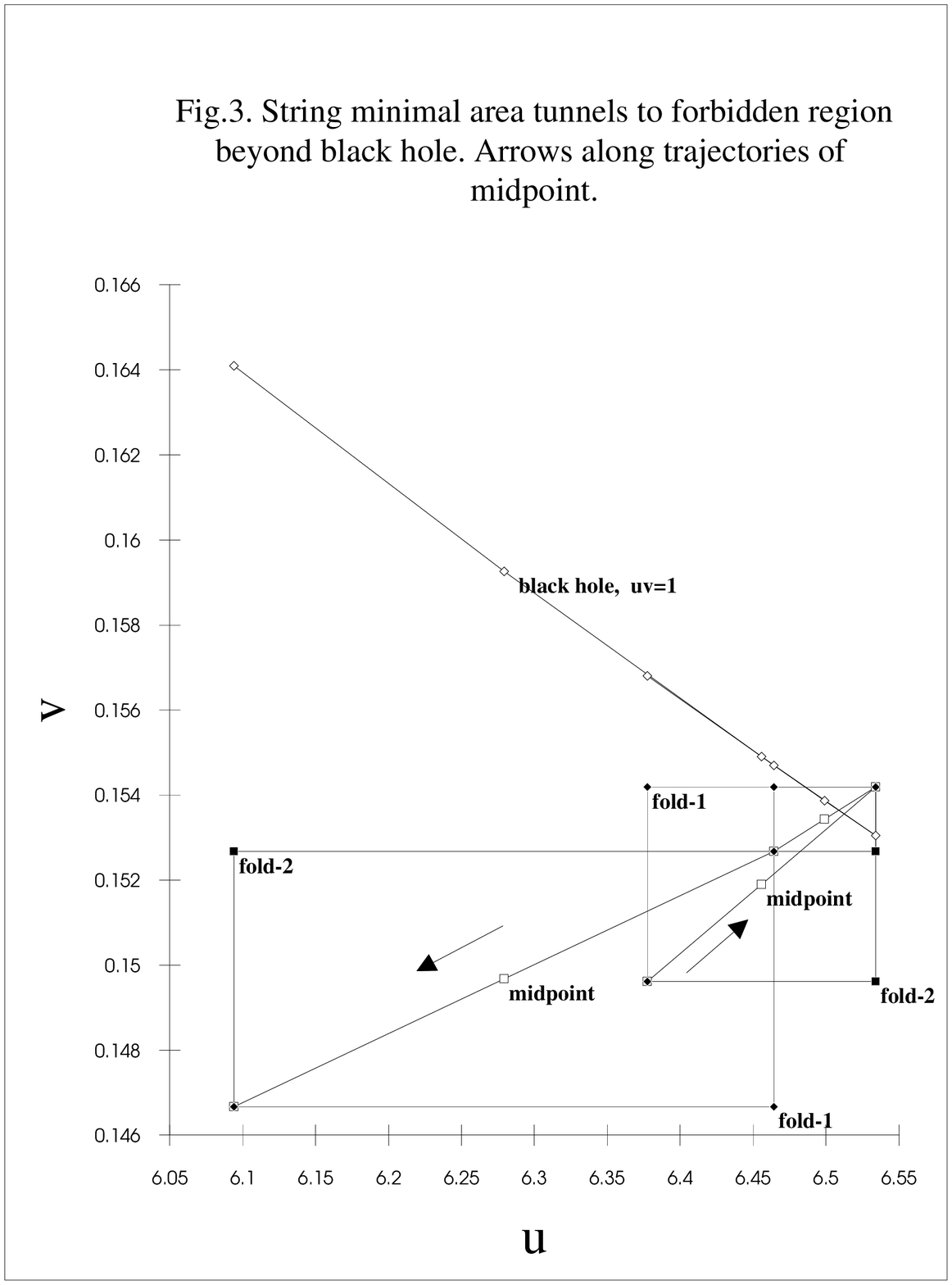}{}{150mm}

\putfig{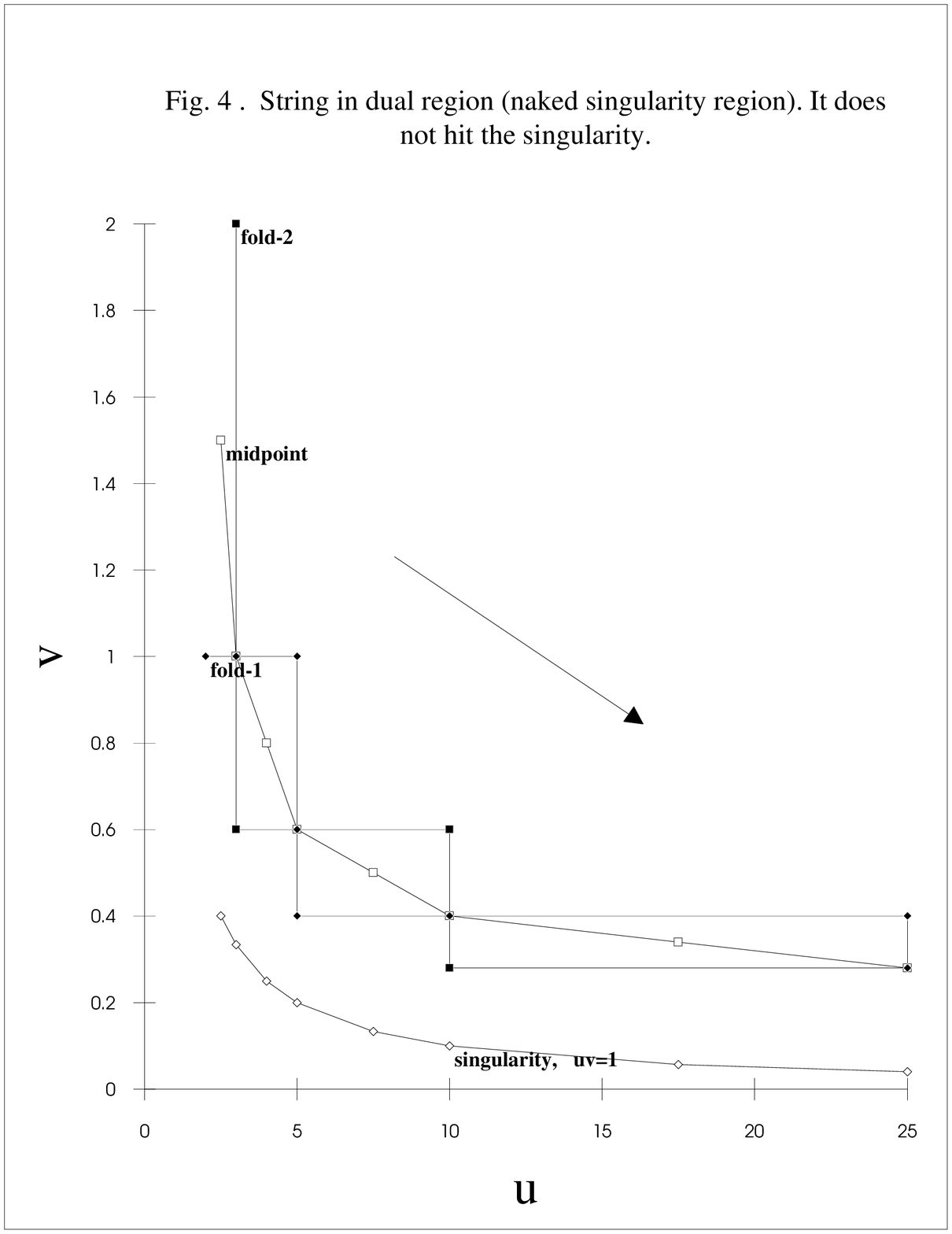}{}{150mm}

\end{document}